\documentclass[final,3p,times,twocolumn]{elsarticle}

\usepackage{epsfig}
\usepackage{lineno}
\usepackage{natbib}

\usepackage{orcidlink}
\usepackage{amsmath,amssymb,amsfonts}

\usepackage{algorithmic}

\usepackage{graphicx}
\usepackage{tabularx}
\usepackage{array}
\usepackage{booktabs}
\usepackage{placeins}
\usepackage{enumitem}
\usepackage{stfloats}

\usepackage{textcomp}

\usepackage{xcolor}
\usepackage{acronym}

\journal{Advanced Engineering Informatics}

\begin{document}

\begin{frontmatter}
\title{SHIA: A Direct SysML–Hardware Interface Architecture for Model-Centric Verification}

\author[inst1,inst2]{Charles Lewis}
\ead{Charles.Lewis@mbda.co.uk}

\author[inst2]{Amal Elsokary}
\ead{a.elsokary@lboro.ac.uk}

\author[inst2]{Siyuan Ji\corref{cor1}}
\ead{S.Ji@lboro.ac.uk}

\cortext[cor1]{Siyuan Ji}

\affiliation[inst1]{
    organization={MBDA},
    addressline={Six Hills way},
    city={Stevenage},
    postcode={SG1 2DA},
    country={UK}
}
\affiliation[inst2]{
    organization={Wolfson School of Mechanical, Electrical and Manufacturing Engineering, Loughborough University},
    addressline={Epinal way},
    city={Loughborough},
    postcode={LE11 3TU},
    country={UK}
}

\begin{abstract}

Model-Based Systems Engineering (MBSE) is widely treated as the backbone of digital engineering, with languages such as the Systems Modeling Language (SysML) providing the means to capture system structure, behaviour, and verification intent. Yet once verification moves to hardware, the system model is routinely left behind. Domain-specific simulation environments, model transformations, and bespoke tool integrations take over, and the model that began as the authoritative reference drifts out of sync with the implementation it was meant to govern. This paper introduces the SysML Hardware Interface Architecture (\ac{SHIA}), which keeps an executable SysML model directly inside the verification loop, exchanging messages with physical hardware without intermediate transformation chains, co-simulation platforms, or broker-mediated plugins. \ac{SHIA} is realised through a SysML side server, written in embedded C++ within IBM Rhapsody, and a hardware side server running on a Raspberry Pi, together establishing a bidirectional link between the digital model and the physical system. A logic gate case study demonstrates the approach end-to-end, from hardware model construction and prototype assembly to test harness design, behavioural statechart control, and staged verification of each component before integration. The integrated system exchanged messages correctly in both directions, and Karnaugh map comparison between the SysML-generated and hardware-generated outputs showed zero discrepancy. The result shows that, when paired with a suitable interface, SysML need not remain a static description that informs downstream tools; it can serve as the executable layer through which hardware behaviour is stimulated, observed, and verified. The work demonstrates a route to model-governed verification and a shorter digital thread between system architecture and the hardware that realises it.

\end{abstract}



\end{frontmatter}

\section*{Abbreviation}
\footnotesize
\begin{acronym}[Systems]
  \acro{FMI}{Functional Mock-up Interface}
  \acro{HiL}{Hardware-in-the-Loop}
  \acro{HiLeS}{High-Level Specification of Embedded Systems}
  \acro{IBD}{Internal Block Diagram}
  \acro{M2M}{Model-to-Model}
  \acro{M2T}{Model-to-Text}
  \acro{MBSE}{Model-Based Systems Engineering}
  \acro{MOM}{Model Only Mode}
  \acro{MRM}{Model Replacement Mode}
  \acro{SHIA}{SysML--Hardware Interface Architecture}
  \acro{SysML}{Systems Modeling Language}
  \acro{SysPhS}{SysML Extension for Physical Interaction and Signal Flow Simulation}
  \acro{VV}[V\&V]{Verification and Validation}
\end{acronym}

\newpage

\section{Introduction}
\label{Intro}

\ac{MBSE} has become the dominant approach to managing the complexity of modern engineered systems, supporting requirements definition, architecture, design, integration, \ac{VV} across the lifecycle through digital models that act as a central reference rather than scattered documentation \cite{ISOStandard,carroll2016systematic,fieber2014assessing,harvey2012document}. The promise is consistency, traceability, and earlier identification of integration risk, with shorter and more predictable development cycles \cite{cederbladh2024early,nolan2008model}. The International Council on Systems Engineering identifies the future of systems engineering as increasingly model-based, enabled by digital transformation, while recent industry surveys and adoption studies show that MBSE is already being taken up across industrial organisations \cite{miller2022future, akundi2022perceptions}.

Within this model-centric foundation, the \ac{SysML} has become the standard for representing system structure, behaviour, requirements, and constraints in a form that supports analysis and verification \cite{de2022taxonomy, hause2006sysml}. Industrial modelling environments such as MagicDraw/Cameo \cite{DassaultSysMLPlugin} and IBM Engineering Rhapsody \cite{IBMRhapsody} make it practical to develop SysML models across multiple levels of abstraction, from high-level system boundaries to detailed internal exchanges \cite{fosse2013systems}. When system behaviour must be assessed against physical implementation, rather than remaining in simulation, \ac{HiL} testing provides the means to exercise hardware against an executable virtual environment, while digital twin concepts extend this idea by maintaining a virtual counterpart linked to the physical system through development and operation \cite{mihalivc2022hardware,tao2019digital,bacic2005hardware, liu2021digital,huang2020towards}.

\begin{figure}[t!]
    \centering
    \includegraphics[width=0.9\linewidth]{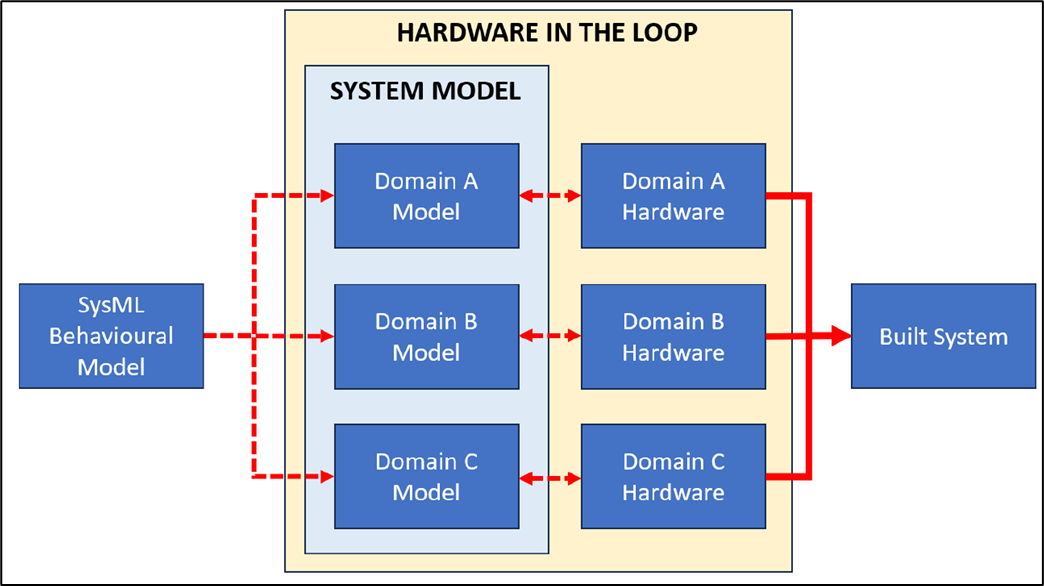}
    \caption{Conventional \ac{HiL} workflow in which SysML informs domain models but remains outside the active verification loop}
    \label{fig:detached}
\end{figure}

Despite the centrality of SysML in design, model-based testing of hardware typically requires coupling SysML with domain-specific tools such as Simulink, Modelica, or \ac{FMI}-based environments, which provide the numerical solvers, physical semantics, and execution capabilities that SysML does not natively support \cite{nigischer2021multi}. The connection between SysML and these tools is described as part of a digital thread: a connected chain of models, simulations, data, and verification evidence intended to maintain continuity across the lifecycle \cite{pessoa2022model}. In practice, however, this chain depends on transformation steps, tool integrations, and manual traceability links that are difficult to keep synchronised as the system evolves \cite{sprock2026bridging}. The result is a recurring problem: SysML captures the system at design time, then is left behind once domain-specific models take over for verification. Changes captured in those downstream models are not reliably propagated back, and the model that was meant to be the authoritative reference becomes outdated, weakening the traceability the digital thread was intended to preserve \cite{kalawsky2013bridging}, as illustrated in Figure~\ref{fig:detached}.


Existing responses to this problem have largely worked within the same paradigm, relying on model transformation chains, co-simulation platforms, or broker-mediated plugins to connect SysML with hardware test environments. Each introduces additional artefacts and intermediary steps that must be configured, maintained, and verified alongside the SysML model, compounding rather than resolving the synchronisation problem. A different approach is needed: one that eliminates the intermediary steps entirely and keeps the SysML model directly inside the verification loop as the authoritative system reference.

\begin{figure}[t!]
    \centering
    \includegraphics[width=0.9\linewidth]{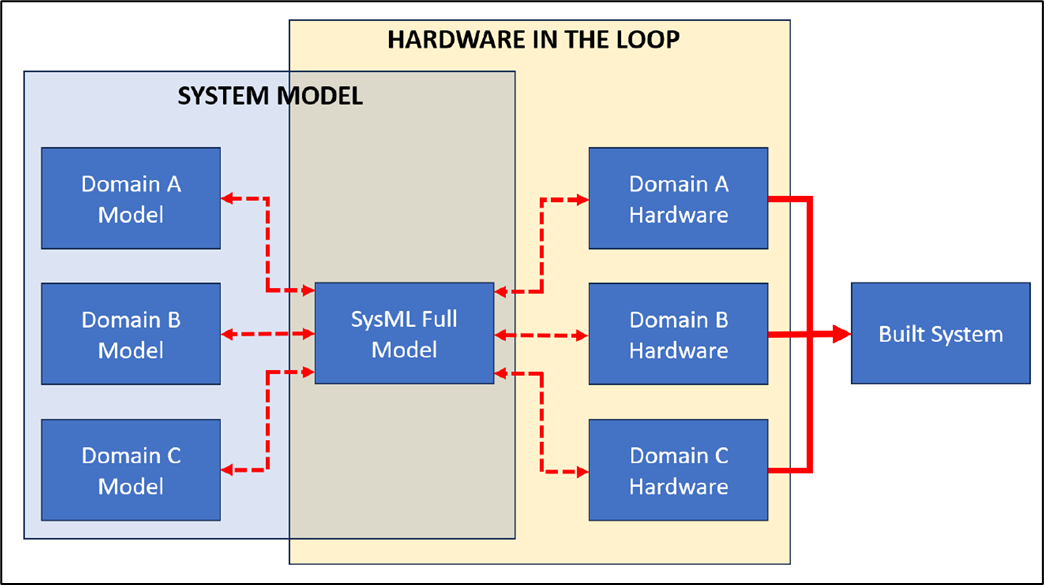}
    \caption{SysML-integrated HiL workflow in which the SysML model remains in the verification loop as the master system model.}
    \label{fig:transition}
\end{figure}
This paper addresses that gap by proposing the \ac{SHIA}, a direct integration approach in which an executable SysML model remains active inside the verification loop and exchanges messages with physical hardware in real time. \ac{SHIA} is distinguished from prior work in three respects. First, it eliminates intermediate transformation chains, co-simulation platforms, and broker-mediated connectors, replacing them with a SysML-side server hosted within the modelling environment and a hardware-side server running on the physical platform. Second, communication is bidirectional and operates without manual intervention, allowing the SysML model to both stimulate the hardware and observe its response within the same authoritative environment. Third, the SysML model functions not as a static design artefact but as an executable verification layer, against which hardware behaviour can be exercised, recorded, and assessed. As shown in Figure~\ref{fig:transition}, this arrangement keeps SysML inside the verification loop as the central system reference, while still permitting domain-specific models to be developed where detailed physical behaviour requires them. The approach is demonstrated through a logic-gate case study covering model construction, prototype implementation, test harness design, behavioural statechart control, and staged verification of the model, hardware, interface mechanism, and integrated system. The contribution is therefore both architectural, in defining a reusable interface mechanism, and methodological, in showing how SysML can be retained as the authoritative reference throughout design, implementation, and verification.


The remainder of this paper is organised as follows. Section \ref{Background} reviews prior work on SysML–hardware integration, examining the model transformation approaches that dominate the literature, surveying related digital engineering and digital twin concepts, and positioning the research contribution. Section \ref{Approach} defines the proposed workflow and implementation requirements. Section \ref{Architecture} presents the \ac{SHIA} design and its physical realisation. Section \ref{caseStudy} describes the case study as proof of concept, including the integration and verification activities. Section \ref{discussion} discusses the key findings and considerations for further development. Section \ref{Conclusion} concludes the paper.

\section{Literature Review}
\label{Background}

Designing, developing, and operating complex systems under a wide range of conditions is inherently challenging. In many cases, it is neither practical nor possible to reproduce all real-world operating conditions during development, yet failures after deployment can result in significant additional cost, time loss, performance degradation, and, in some cases, irreversible damage. For this reason, early \ac{VV} are essential to reduce risk, identify issues before implementation or operation, and build confidence in system behaviour under both expected and unforeseen conditions. This importance is widely recognised in the literature, leading to proposals of a range of \ac{VV} approaches. Among these, this section focuses on \ac{HiL} as an important early verification method for complex systems \cite{kiesbye2019hardware}. \ac{HiL} enables physical hardware to be tested within a real-time simulated environment, providing a greater level of realism for integration and implementation testing.

Recent use of the \ac{HiL} terminology in explicit connection with SysML is evident in two 2024 studies. Yeiser et al.\ \cite{yeiser2024exploring} explored executable SysML in CATIA Magic through interaction with a LEGO Mindstorms prototype, demonstrating that the model could receive hardware inputs and issue control actions, but also reporting that the available hardware interface functions did not permit full integration or in-depth closed-loop testing; in particular, limitations in sensor-mode switching and runtime motor-angle reading prevented real-time behaviour such as target tracking. Helle et al.\ \cite{helle2024hardware} presented a Cameo-based \ac{HiL} approach in which an MQTT Simulation Connector plugin enabled bidirectional communication between SysML simulations and hardware via an MQTT broker, thereby extending SysML execution into a hardware test setting. However, the approach is realised through an external plugin and broker-mediated communication architecture, and is positioned for Cameo-based simulations rather than as a direct native hardware-driving mechanism within the SysML model itself.

Taken together, these studies show that SysML-based \ac{HiL} is beginning to emerge in the literature, but current implementations remain either exploratory and tool-limited, or dependent on external connector infrastructures rather than demonstrating a fully direct, reusable, model-centred hardware interface. Published research on the use of SysML to directly drive hardware, therefore, remains sparse, suggesting that the concept is still relatively novel. The wider literature has instead focused primarily on integrating SysML with external models and specialised toolchains in order to enable interaction with hardware. Within this broader body of work, three recurring themes emerge: model transformation, \ac{HiLeS}, and digital engineering.

\subsection{\textbf{Model Transformation}}

\begin{figure}[t!]
    \centering
    \includegraphics[width=1\linewidth]{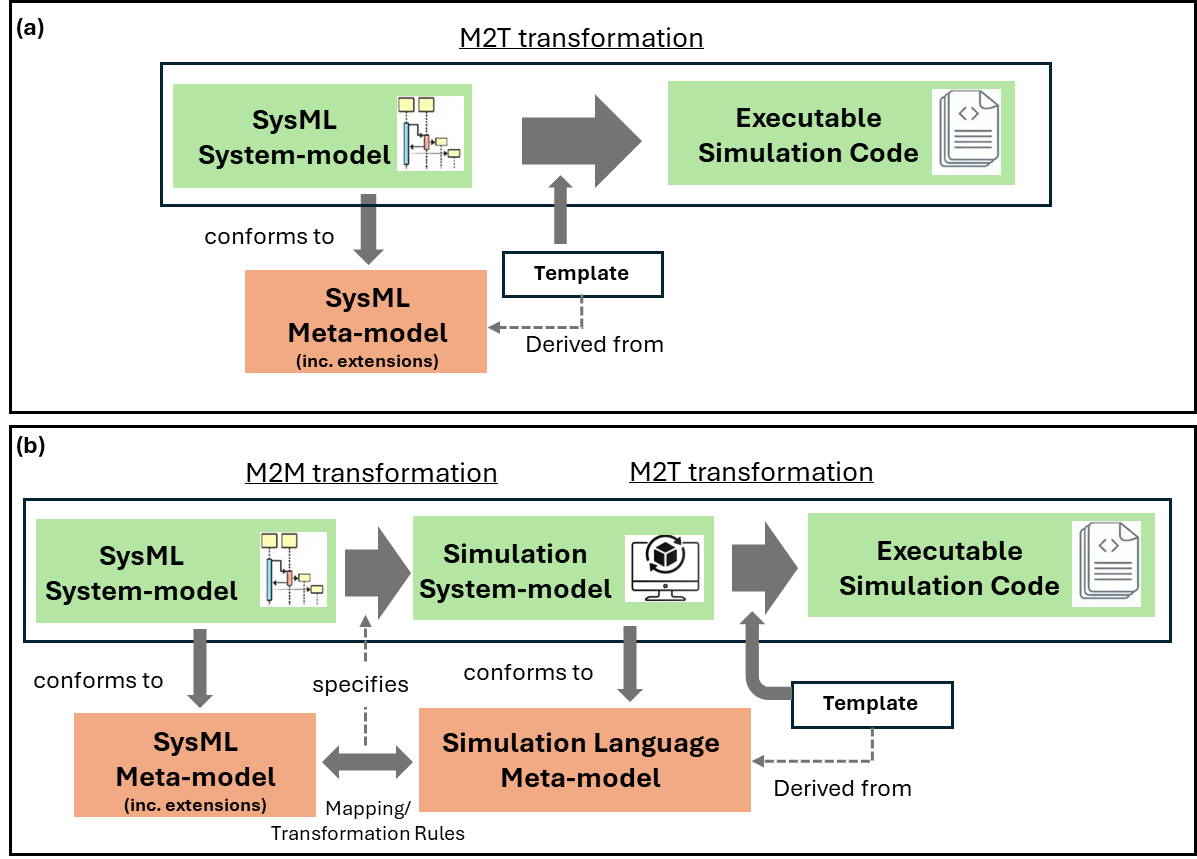}
    \caption{(a \& b) \ac{M2T} vs. \ac{M2M} transformation approaches}
    \label{fig:transformation}
\end{figure}

The vast majority of papers concerned with the practical use of SysML outputs rely on \ac{M2M} and \ac{M2T} transformations, introducing an additional step between the SysML model and the eventual hardware-integration environment, often through a domain-specific platform. Figure~\ref{fig:transformation} (a,b) compares two approaches for generating executable simulation code from a SysML system model. In the upper approach, the SysML model conforms to the SysML meta-model and is transformed directly into executable simulation code through an \ac{M2T} transformation using templates derived from the meta-model. In the lower approach, the transformation is performed in two stages. First, the SysML model is converted into a simulation system model through an \ac{M2M} transformation based on mapping rules between the SysML meta-model and the simulation language meta-model. The intermediate \ac{M2M} step helps resolve the semantic mismatch between SysML and the target simulation language. The resulting simulation model, which conforms to the simulation language meta-model, is then transformed into executable simulation code through a second model-to-text transformation. In such workflows, SysML typically remains the system-level modelling environment, while platforms such as Speedgoat \cite{speedgoat2025solutions} provide the real-time execution platform for control designs and plant simulations. 

To support these downstream transformations, the literature also reports the use of various plugins, profiles, and modelling constraints within the SysML environment to improve compatibility and compliance with the target domain-tool workflow \cite{nigischer2021multi}. In this category, there are two representative approaches: the SysML4Simulink profile \cite{chabibi2018towards} and the \ac{SysPhS} \cite{Jankevicius2020SysPhS}.

The SysML4Simulink profile provides a structured method for \ac{M2T} conversion using the Acceleo tool, transforming SysML \acp{IBD} and parametric diagrams into MATLAB scripts executable in Simulink \cite{chabibi2018towards}. Although this method demonstrates the feasibility of SysML model conversion, it lacks real-time capability and depends on an additional model, the MATLAB script, undermining the single-master-model concept.  

Similarly, \ac{SysPhS} defines an OMG standard extension that allows a SysML model to be transformed into Modelica or Simulink/Simscape formats \cite{omg2021sysphs}. While this enhances SysML’s ability to represent physical interactions, it also relies on \ac{M2M} transformation and therefore prevents real-time hardware integration. However, if direct real-time SysML–hardware interfacing were achieved, \ac{SysPhS} could substantially extend SysML’s capability to model complex systems down to the component level \cite{barbau2019translator}.

Beyond these two, several other transformation approaches are explored in the literature. The \ac{FMI} standard provides a well-documented framework for exchanging model data via Functional Mock-up Units (FMUs) containing XML, binary, and C code \cite{fmi2023spec}. Although flexible and model-agnostic, it requires models to be converted into FMUs, which limits its suitability for direct real-time SysML–hardware integration. Tools such as HybridSim \cite{wang2013hybridsim} enable SysML-to-FMU conversion but still necessitate scripting via the \ac{FMI}-API to communicate with hardware.

Furthermore, a SysML-to-Simulink transformation approach based on XSLT is also discussed to address the limitation that existing tools do not provide full end-to-end support for the complete \ac{MBSE} process by proposing a bidirectional \ac{M2M} transformation framework \cite{kalawsky2013bridging}. In this approach, SysML models are exported in XMI format and translated into Simulink MDL files using Extensible Stylesheet Language Transformations (XSLT). Although the method is reported to be reliable, it does not support real-time integration and still depends on transformation templates that must be configured, maintained, and verified. As a result, the approach introduces multiple model representations, which can weaken the single-source-of-truth principle and increase configuration-management effort.

In a related context, a real-time communication in control-system applications is demonstrated by a Speedgoat-based platform, using User Datagram Protocol (UDP) and the Experimental Physics and Industrial Control System (EPICS) \cite{rossa2024epics}. Although the work is not specific to SysML, it highlights a possible route for real-time hardware communication. However, such an approach would involve greater architectural complexity and more demanding hardware requirements.

Finally, two additional studies illustrate how SysML transformation principles can be extended beyond conventional model exchange. One proposes SysML-to-software generation for real-time robotic control by automatically translating XML-based SysML state charts into C code \cite{godart2017generating}. Another, the Virtual Factory (VF) environment, retains a master SysML model that is verified against physical test results through a bespoke integrated data environment linking MagicDraw, DOORS, and Manufacturing Execution Systems (MES) \cite{wang2019model}. Together, these studies show that SysML-based transformation and integration can support both executable control implementation and tighter connection between system models and physical validation.

\subsection{\textbf{\ac{HiLeS}}}

A recurring theme in the literature on SysML–hardware integration is the \ac{HiLeS} framework, developed at the University of Los Andes, Colombia. \ac{HiLeS} interfaces between SysML, Petri nets, and VHDL, serving as a modelling tool for embedded systems design validation \cite{gomez2010embedded}. Similar to SysML in its block-based representation of structures, activities, and sequences, \ac{HiLeS} distinguishes itself through \ac{M2M} transformations into Petri nets and VHDL, enabling formal verification and simulation-based validation. SysML integration was introduced in \ac{HiLeS}2, which, as demonstrated by Hoyos et al.\ \cite{hoyos2011hiles2}, provides a method for developing system-level SysML models that support rapid prototyping and automatic generation of deployable VHDL firmware.

\ac{HiLeS} enhances model verification by transforming SysML sequence diagrams into Petri nets and validating the resulting VHDL simulations. Petri nets represent information flow through tokens that move between states and transitions, effectively capturing concurrent and parallel activities that are often difficult to express in SysML activity or state diagrams. Their strong mathematical foundation enables formal and rigorous analysis, ensuring robust verification of system logic and flow.

Further papers related to \ac{HiLeS} provide additional case studies demonstrating the framework’s ability to transform SysML models into Petri nets \cite{gutierrez2014hardware,gutierrez2015hardware}. The former paper is mathematically heavy, and the focus is naturally on the eventual generation of embedded VHDL code from the SysML starting point. The latter is more systems-focused, but the end goal remains the generation of VHDL code, and so in the context of this project, \ac{HiLeS} can be summarised as another \ac{M2T} transformation tool. However, the concept of the verification of SysML sequence diagrams by transforming them into Petri nets is an interesting one, and could be considered as a method to provide additional verification to a system or subsystem SysML model. In summary, \ac{HiLeS} is limited in its suitability for system-level development because it operates at a relatively low level of abstraction; nevertheless, its verification mechanism remains a promising concept for future applications.

\subsection{\textbf{Digital Engineering}}\

Driven by cost and performance pressures, the engineering industry is increasingly transitioning design and testing activities from the physical to the virtual domains. As the focus of this project is on SysML and \ac{MBSE}, this makes it well within the realms of digital engineering. In this context, Güdemann et al. \cite{gudemann2010sysml} demonstrated the application of SysML in modelling an air cargo hub, where well-defined interfaces and data enabled subsystem simulations to identify system-level weaknesses before physical testing. The subsequent development of SysML v2 extends these capabilities through an open API and improved metamodel structure, promoting interoperability across digital tools and enhancing model-based \ac{VV} \cite{bajaj2022systems}.

Building on this foundation, once a verified SysML subsystem model is realised in hardware, it effectively becomes its Digital Twin, a virtual counterpart that mirrors and interacts with the physical system. This aligns with the characterisation of the Digital Twin concept by Jones et al. \cite{jones2020characterising}, who describe progressive phases from virtual entity creation through metrology and realisation, where feedback from the physical system continually refines the virtual model’s fidelity. Similar principles appear in Grieves's paper \cite{grieves2022digital}, whose “virtual-to-real factory” model highlights the importance of a Unified Repository linking design data, system attributes, and sensor feedback to enable real-time synchronisation between virtual and physical entities.


A large-scale application of this concept was demonstrated in NASA’s Orion spacecraft Digital Twin project, which used an executable SysML model as the system’s digital core \cite{pierce2023orion}. Although challenged by the late introduction of digital modelling and the absence of real-time metrology, the project successfully consolidated distributed design data into a single repository and automated several analysis processes. Building on these insights, the proposed SysML–Hardware interface integrates twinning principles from the outset, enabling parameter monitoring and bidirectional data exchange. This approach not only addresses earlier limitations but also aligns with broader Industry 4.0 developments, where IoT, AI, and machine learning drive continuous synchronisation, comparison, and optimisation between virtual models and their real-world counterparts.

\subsection{Summary}

\begin{figure}[ht!]
    \centering
    \includegraphics[width=1\linewidth]{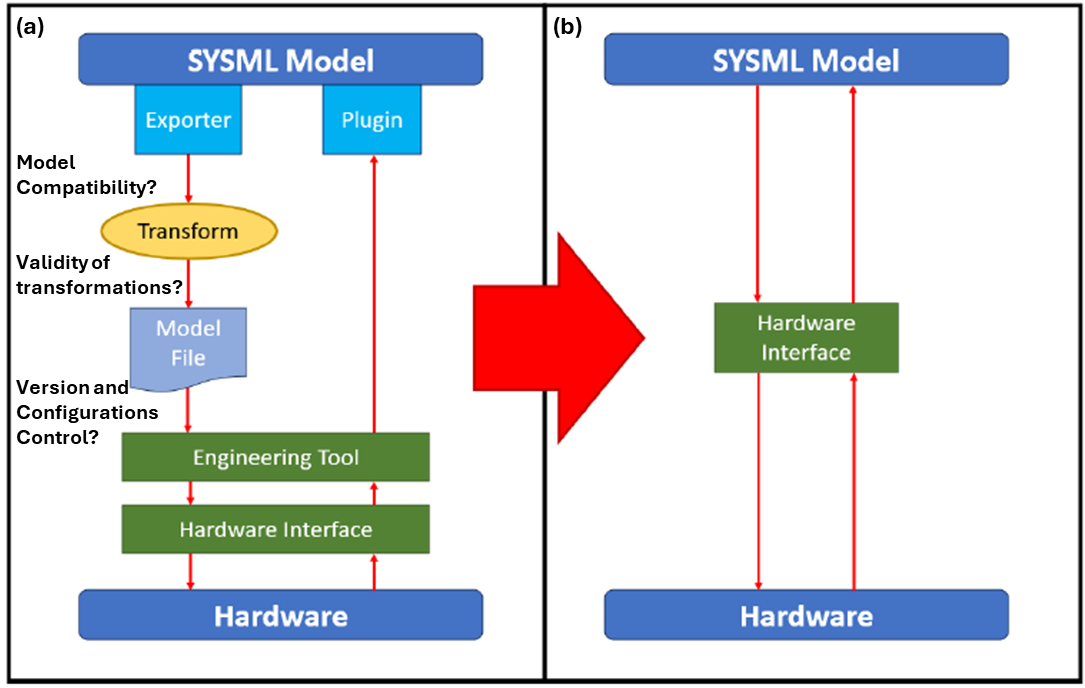}
    \caption{(a) General considerations required for model transformation and hardware interfacing, (b) Distillation of model transform interface to ideal interface}
    \label{fig:Model}
\end{figure}

In summary, the reviewed research illustrates the breadth of SysML transformation approaches, each contributing valuable techniques for co-simulation, code generation, and validation. However, they share a common limitation: the numerous intermediary steps required to connect SysML models with hardware. As shown in Figure~\ref{fig:Model} (a), each transformation stage introduces risks to process validity and consistency, as well as considerable engineering overheads through the configuration and maintenance of multiple tools and plugins.

Based on the preceding review of the literature, the aim of this research is to establish a direct SysML–hardware interface that eliminates these intermediary steps, enabling seamless, automated and verifiable communication between the virtual and physical counterparts. As illustrated in Figure~\ref{fig:Model} (b), this project proposes an ideal distillation of the generic model transformation process, reducing it to a single controllable element. By applying strict configuration management only at this interface, dependency on external tools is minimised, while integrity between SysML models and their corresponding hardware systems is maintained.

\section{Proposed Approach}
\label{Approach}

Through this section, a direct interfacing mechanism is developed between a system model described using SysML (referred to as the SysML model for simplicity) and physical hardware. The proposed interfacing mechanism, \ac{SHIA}, is intended to minimise the number of intermediary steps between the SysML model and the hardware, thereby reducing the need for additional interim verification stages. \ac{SHIA} is developed to support the automatic exchange of messages between the SysML model and hardware without user intervention, enabling a continuous and real-time flow of information. This essentially makes the SysML model a true digital twin of the hardware that it is interfacing with. 

Furthermore, \ac{SHIA} interfacing mechanism is designed to allow SysML to remain the single source of truth throughout design, implementation, and verification. Because the system design originates in SysML, the model is retained as the authoritative system-level representation as the hardware evolves, keeping SysML actively involved in hardware verification. This strengthens traceability, reduces model–hardware divergence, and improves consistency between system architecture, hardware realisation, and verification evidence. 

\begin{figure*}[ht]
    \centering
    \includegraphics[width=1\linewidth]{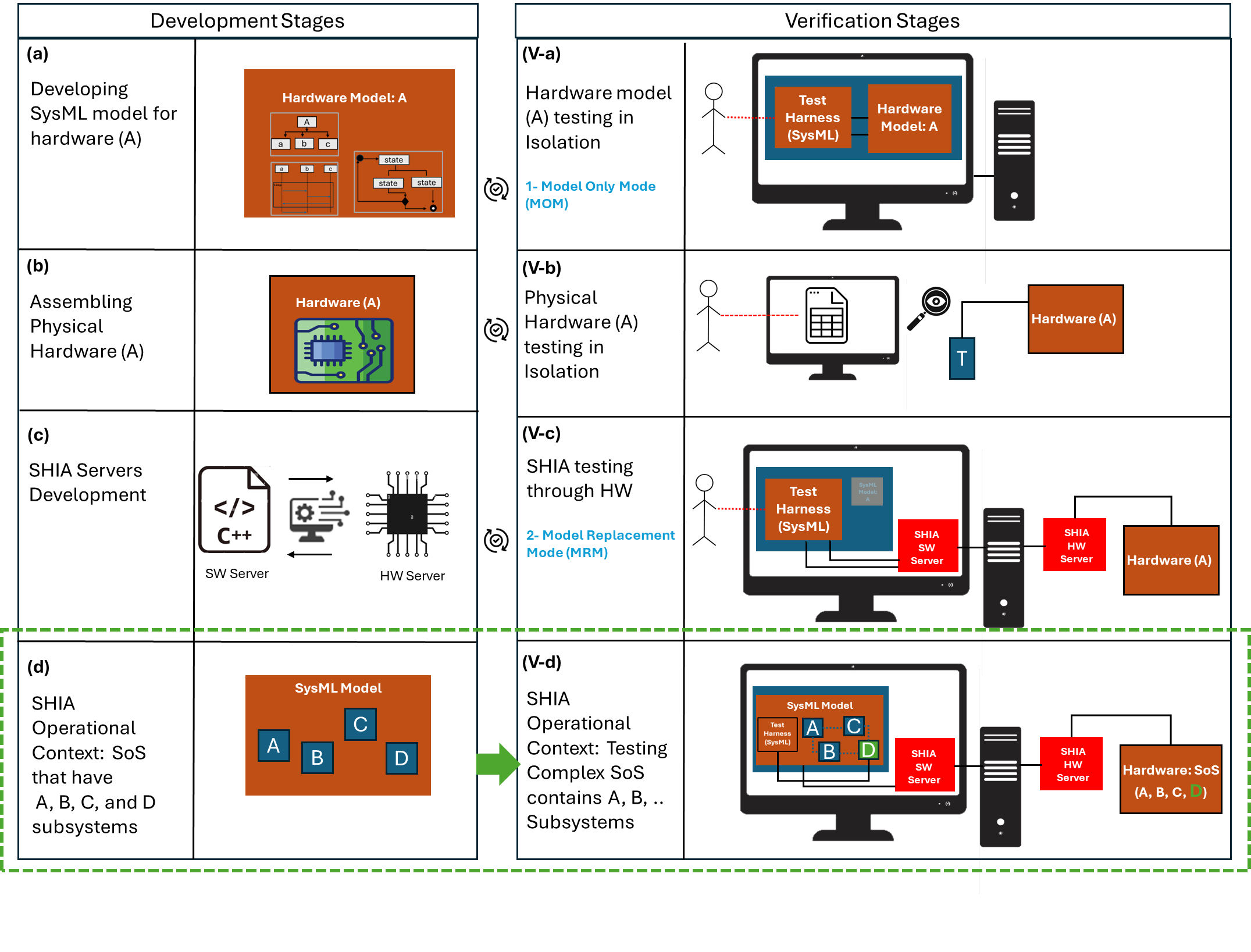}
    \caption{Development and verification stages of the proposed SHIA workflow. (a) shows SysML modelling, (b) hardware assembly, (c) SHIA servers implementation, and (d) system-of-systems operation. each step is paired with a corresponding verification stage for iterative refinement.}
    \label{fig:prop}
\end{figure*}

To achieve this aim, a repeatable development and verification workflow was proposed for implementing a generic interface mechanism. As shown in Figure~\ref{fig:prop}, the workflow is organised into paired development and verification stages. The development stages describe the progressive construction of the SysML hardware model, the physical hardware, the SHIA software and hardware servers, and finally the operational system-of-systems context. Each development step is then linked to a corresponding verification activity, in which the model, hardware, or interface is tested before progressing to the next stage. In this way, the workflow supports iterative refinement: the outcomes of each verification stage inform the improvement of the corresponding development stage. This provides both a structured guide for the case study implementation presented in Section~\ref{caseStudy} and a general approach for developing and verifying hardware directly through its SysML model.

\subsection{Step 1: Develop SysML model for the hardware:}

The first step involves creating a set of SysML diagrams in sufficient detail to represent the structure and operation of the intended hardware to be built and tested. This requires the model to be refined beyond a high-level abstraction to the level of lower-level building blocks, with explicit representation of components, interfaces, interactions, interconnections, and behaviours, for example, through \ac{IBD}s and state charts. Therefore, the resulting model faithfully mirrors the hardware intended for implementation and can serve as an executable representation of the system, as shown in Figure~\ref{fig:prop} (a).

\subsection{Step 2: Build a prototype hardware (A):}

Using the SysML model created in the first step as the basis, the second step is to construct a prototype that acts as the hardware (HW) counterpart of the developed SysML model. Figure~\ref{fig:prop} (b) shows the assembling of a prototype that serves as the physical twin of the model for later verification and integration. During the verification process, both the hardware and the SysML model can be refined iteratively: updates to the model can be implemented in the hardware, and changes in the hardware assembly, configuration, or observed behaviour can be reflected back into the SysML model. This helps maintain consistency between the physical system and its model, while enabling the model to capture the system’s behavioural performance.

\subsection{Step 3: Develop the SysML--Hardware Interface (SHIA):}

Figure ~\ref{fig:prop} (c) includes the development of SHIA servers:
\begin{enumerate}
\item preparation of the SysML-side interface; and
\item preparation the hardware-side interface.
\end{enumerate}

Because SysML does not provide native support for direct real-time communication with a computer port, part of the \ac{SHIA} system must be implemented in a software counterpart. At the same time, a hardware element is required to enable communication with the developed prototype and other generic hardware devices. This leads to the identification of two functional ports for the  \ac{SHIA} mechanism: a software server and a hardware server. On the SysML side, \ac{SHIA} must receive and transmit data, while on the hardware side, it must set and read signals.

\subsection{Step 4: Verify System Elements}

This verification step is organised into four stages, corresponding to each development step, as illustrated in the right hand side of Figure~\ref{fig:prop}, in order to establish confidence in the individual elements, and subsequently, in their integrated operation. Central to this process is the test harness, which is developed within the SysML environment to provide both the stimulus source and the observation interface for verification activities. Through this arrangement, the operator can apply inputs, monitor outputs, and assess system responses under controlled conditions. The test harness supports two principal modes of operation that will be discussed in subsequent sections.

\subsubsection{Verification of the SysML model in isolation}
\label{sysMLTest}

The SysML model is initially verified independently of any physical hardware by using a test harness that applies input stimuli and records the corresponding outputs in a truth table. As such, this stage is referred to as a \ac{MOM} verification, in which the SysML model of the intended hardware is tested in isolation so that its behavioural and logical performance can be confirmed. \ac{MOM} is shown in Figure~\ref{fig:prop} (V--a). By applying all required input combinations through the developed SysML diagrams and observing the simulated responses, the logical and behavioural correctness of the model can be verified. 

\subsubsection{Verification of the prototype in isolation} 
\label{HWTest}

Following verification of the SysML model, the hardware prototype is tested independently by applying direct stimuli to its physical inputs and recording the corresponding outputs. This stage is intended to establish confidence that the hardware reproduces the intended model behaviour, as shown in Figure~\ref{fig:prop} (V--b). Once this has been confirmed, the same standalone manual verification would not normally need to be repeated for later equivalent hardware, since its purpose here is to provide confidence that both the model side and the hardware side are functioning correctly before \ac{SHIA} is introduced.

\subsubsection{Verification of the \ac{SHIA} mechanism}
\label{SHIATest}

This stage focuses on verifying the operation of the \ac{SHIA} mechanism by connecting it to the already verified hardware, allowing the hardware output responses to be captured and transmitted to the test harness, which remains the same stimulus environment previously used for the SysML model. To enable this configuration, \ac{MRM} is adopted within the test harness to support verification of \ac{SHIA} operability while also demonstrating that subsequent hardware verification can be performed through the same setup. In this way, the arrangement confirms the reliable operation of the \ac{SHIA} communication mechanism and enables future verification of the hardware itself by capturing its behaviour in truth tables and comparing its performance with the previously recorded truth tables of its SysML model. This supports cyber-physical integration and real-time verification, as illustrated in Figure~\ref{fig:prop}(V--c)

\subsubsection{Verification of the integrated system:}

Once the individual verification activities are complete in the previous stages, the integrated system is verified as a whole to ensure correct communication and behavioural consistency between the SysML model, \ac{SHIA}, and the hardware prototype. This stage demonstrates that changes in one domain can be reflected in the other, thereby confirming the effectiveness of the interface in supporting virtual verification of HW in future iterations.

In addition, it provides a basis for discussing the operability and potential reusability of \ac{SHIA} as a general mechanism for SysML--hardware integration. An important operational use case for a mature \ac{SHIA} implementation is the assessment of alternative or replacement subsystems, such as those introduced because of obsolescence. By connecting the physical subsystem to the wider system model through \ac{SHIA}, while bypassing the original modelled subsystem, its effect on the overall system can be evaluated efficiently, as shown in Figure~\ref{fig:prop} (d) and (V--d).

\subsection{Step 5: Proof of Concept}

The final step is to prove the concept by applying a case study, recording the workflow, setup steps, and key design considerations required to establish an implementation of the validated interface for future reuse and extension.

\section{SHIA Design \& Implementation}
\label{Architecture}

This section presents the architectural composition of \ac{SHIA} and the main design decisions behind its physical implementation. The architectural design choices follow a technology-agnostic principle, while the physical implementation of \ac{SHIA} requires the selection of specific technologies.

Overall, \ac{SHIA} is divided into a SysML-side interface and a hardware-side interface, which together form a bidirectional real-time bridge between the executable SysML model and the physical hardware without requiring model transformation into an external simulation environment.

\subsection{SysML Side Interface}

The SysML-side interface was designed solely to support communication to and from the executable SysML model. For this reason, it was considered most appropriate to host this interface within the SysML environment itself. Two specific technologies are selected for \ac{SHIA} implementation, namely Rhapsody (which is used for SysML modelling) and C++, which is the programming language that could be inserted directly into the SysML model created in Rhapsody and executed in response to stage changes or transitions. It is important to note that \ac{SHIA} conceptual design is, in principle, agnostic to the modelling and programming technology choices. However, the implementation complexity can vary. In this work, the term \textit{SysML-side server} refers to the operational C++ server implemented within the SysML environment, which is responsible for converting model-generated messages and transmitting them from Rhapsody to physical hardware.

\subsection{Hardware Side Interface}

For this tool-specific implementation, a Raspberry Pi was selected at the hardware end of \ac{SHIA} because it allowed rapid modification during testing. However, for deployment in a production environment, a Field-Programmable Gate Array (FPGA)-based platform would likely be more appropriate for further development because such platforms can provide higher-performance real-time control and lower-latency hardware interaction. One example is National Instruments CompactRIO (cRIO), which combines a controller running NI Linux Real-Time with a user-programmable FPGA. This type of architecture could support a \ac{SHIA}-like implementation while enabling faster module-level interaction through FPGA-based timing and control rather than reliance on a conventional software-managed bus \cite{ni_compactrio,ni_crio_arch,ni_linux_rt}.

\subsection{Communication Protocol}
\label{Comm}

For communication between the SysML model host PC software and the Raspberry Pi, Ethernet and Serial were evaluated as candidate protocols. Their comparison is summarised in Table~\ref{tab:protocol_comparison}, using the criteria of software consistency, hardware requirements, bandwidth, latency, implementation effort, and scalability. As reflected in the table, Ethernet offered advantages in terms of built-in hardware support on the Raspberry Pi 3B through its RJ45 port and in its higher communication bandwidth, whereas Serial provided lower reported latency in a comparable peer-to-peer digital communication study and a simpler implementation route for rapid prototyping \cite{gugerty2006case}.

\begin{table*}[ht]
\centering
\caption{Comparison of candidate communication protocols for \ac{SHIA} implementation}
\label{tab:protocol_comparison}
\renewcommand{\arraystretch}{1.25}
\setlength{\tabcolsep}{5pt}
\begin{tabular}{|p{2.2cm}|p{6.5cm}|p{6.5cm}|}
\hline
\textbf{Criterion} & \textbf{Ethernet} & \textbf{Serial} \\
\hline

Appearance &
\centering\includegraphics[width=0.18\linewidth]{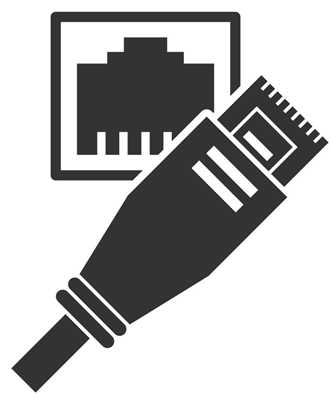}
&
\centering\includegraphics[width=0.2\linewidth]{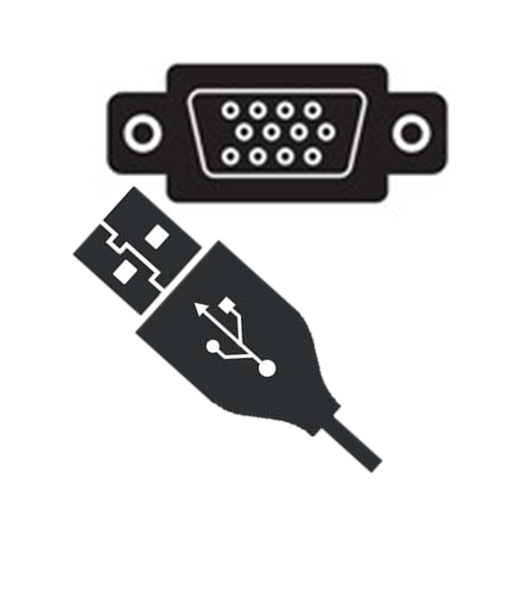}
\tabularnewline
\hline

Hardware &
The Raspberry Pi 3B includes an RJ45 Ethernet port, making physical connection straightforward with minimal additional hardware. &
Serial communication can be achieved through the Raspberry Pi UART on GPIO 14/15, but requires suitable serial interfacing and logic-level compatibility. \\
\hline

Software &
Ethernet allows a straightforward alignment between both sides of the interface through a C++ implementation consistent with the Rhapsody-based SysML server. &
Serial communication leads to a GPIO-supported interface on the Raspberry Pi, which is implemented in C++ to interpret incoming commands and translate them into hardware actions.\\
\hline

Bandwidth &
Ethernet provides substantially higher bandwidth and is well suited to future large-scale or data-intensive integration. &
Serial communication provides lower bandwidth, but this is generally adequate for small command and status messages. \\
\hline

Latency &
In a comparable reported test scenario, RSRP over Ethernet LAN showed a peer-to-peer latency of 14.6 ms. &
In the same study, RSRP over Serial LAN showed a lower peer-to-peer latency of 5.2 ms. \\
\hline

Implementation &
Implementation on Windows typically requires socket-based networking using Winsock, which introduces additional software complexity. &
Serial communication can usually be implemented more directly through standard UART/serial libraries. \\
\hline

Scalability &
Well suited to networked, distributed, and future extensible architectures. &
Better suited to direct point-to-point communication and small-scale integration. \\
\hline

Overall assessment &
Better for long-term scalability and bandwidth. &
Better for low-latency, simple, rapid prototyping. \\
\hline
\end{tabular}
\end{table*}


Both Ethernet and serial communication are suitable for this type of integration. Ethernet was initially considered because it can provide greater scalability and wider bandwidth, which may be beneficial for larger or more complex systems. However, for this proof-of-concept, serial communication was selected because it was sufficient to support the required interaction between the software and hardware components of \ac{SHIA}, and it was also more compatible with the available computer and development setup. Therefore, serial communication provided a practical and effective solution for the current implementation. Future developments, or different system configurations, may adopt Ethernet where its capabilities and compatibility better match the project requirements.

\subsection{Implementation}

\begin{figure}[ht]
    \centering
    \includegraphics[width=1\linewidth]{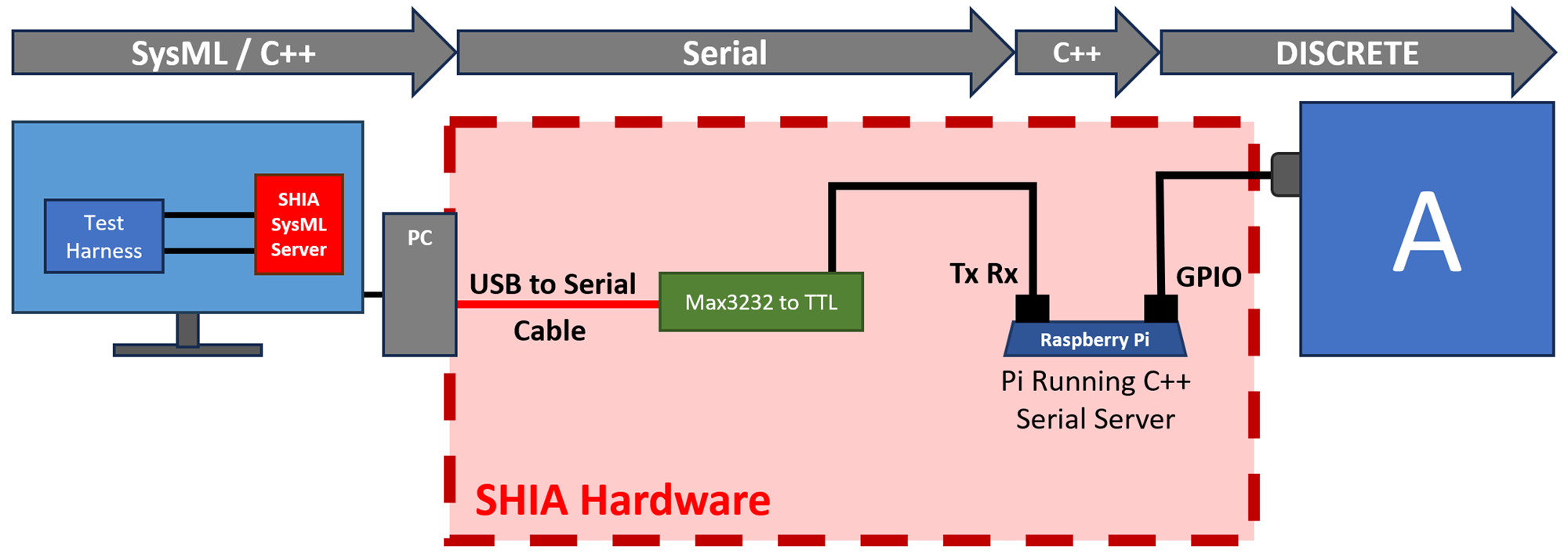}
    \caption{\ac{SHIA} architecture showing the SysML server, serial communication path, and hardware-side SHIA boundary}
    \label{fig:ArchitecSHIA}
\end{figure}

Figure~\ref{fig:ArchitecSHIA} illustrates the \ac{SHIA} communication architecture as a layered end-to-end interface between the executable SysML model and the physical hardware. The architecture comprises two principal elements: a software-side interface hosted on the PC and a hardware-side interface highlighted by the red dashed region, comprising the serial cable, converter, programmed Raspberry Pi, and some kind of interface for the hardware to the GPIO array. To understand the process, on the software side, the SysML test harness and \ac{SHIA} SysML server operate within the Rhapsody/C++ environment to generate and transmit system-level commands.

On the hardware side, these commands are conveyed from the PC through a serial communication link, implemented physically by a USB-to-serial cable. Since the Raspberry Pi’s GPIO pins operate using low-voltage TTL signalling rather than standard serial voltage levels, the communication passes through a MAX3232 serial-to-TTL converter, which safely converts the signal before it reaches the Raspberry Pi. To control the Raspberry Pi hardware interface through the GPIO pins, the open-source WiringPi library was used~\cite{wiringpi_github}. This enables the converted serial commands to be interpreted and mapped directly to the GPIO pins, where they are translated into discrete actions to drive the physical hardware, while the resulting hardware state is returned to the SysML environment.

The overall system therefore comprises a model-side SysML/C++ layer, a serial communication layer, a Raspberry Pi-based C++ control layer, and the discrete physical hardware, which collectively establish a bidirectional real-time bridge between the virtual model and the physical system.

\section{Case Study}
\label{caseStudy}

As discussed in Section~\ref{Approach}, the proposed steps provide a traceable guide for the work presented throughout this section, in which a case study is used to demonstrate the approach and the validity of the \ac{SHIA} design and implementation. For this purpose, a logic-gate arrangement was chosen as the example system. This choice was made for simplicity and practicality, as logic gates can be modelled easily in SysML and implemented in hardware on a prototyping board using low-cost, readily available components. Although a more complex system-of-systems example could have better illustrated \ac{SHIA}’s ability to handle complexity, a simple proof of concept was considered more appropriate for this stage, particularly in light of the limited examples identified in the literature.

The development of these elements requires a multi-stage verification process to ensure that each part is assessed independently before progressing to full system integration. For clarity, the development and isolated verification of each element are presented together within the same subsection. The following terms are also defined in order to ensure consistent reference to the principal elements throughout the section.

\begin{itemize}
    \item \textbf{Hardware model}: the digital representation of the intended hardware within the SysML environment, comprising the structural and behavioural SysML diagrams used to define its design and operation.
     \item \textbf{Physical hardware}: the assembled prototype representing the intended system of interest in its physical form.
    \item \textbf{Test harness}: a dedicated view within the SysML environment that serves two principal functions, depending on the selected mode of operation: \ac{MOM} or \ac{MRM}. This test harness serves as the SysML-side testing point for hardware model or for physical hardware via \ac{SHIA}, through which signals can be monitored.
    \item \textbf{SysML model}: This term is the combination of hardware model and test harness in the sysML environment.
   
\end{itemize}

\subsection{Hardware Model Development}

The \ac{SHIA} process begins with the development of a hardware model before it is linked to its physical hardware counterpart. In this work, the model was developed following a bottom-up approach. Once the example system had been selected, logic gates in this case study, the first modelling task was to develop a suite of reusable logic-gate blocks within the IBM Rhapsody SysML model. An assortment of logic blocks were implemented, one of which being the NAND gate, whose logical behaviour is defined by the relationship between its two inputs and single output. The NAND gate output is given by eqn ~\ref{eq:nand}, which is equal to 1 (High) for all input combinations except when both inputs are 1.

\begin{equation}
Y=\overline{A \cdot B}
\label{eq:nand}
\end{equation}

\begin{figure}[ht]
    \centering
    \includegraphics[width=1\linewidth]{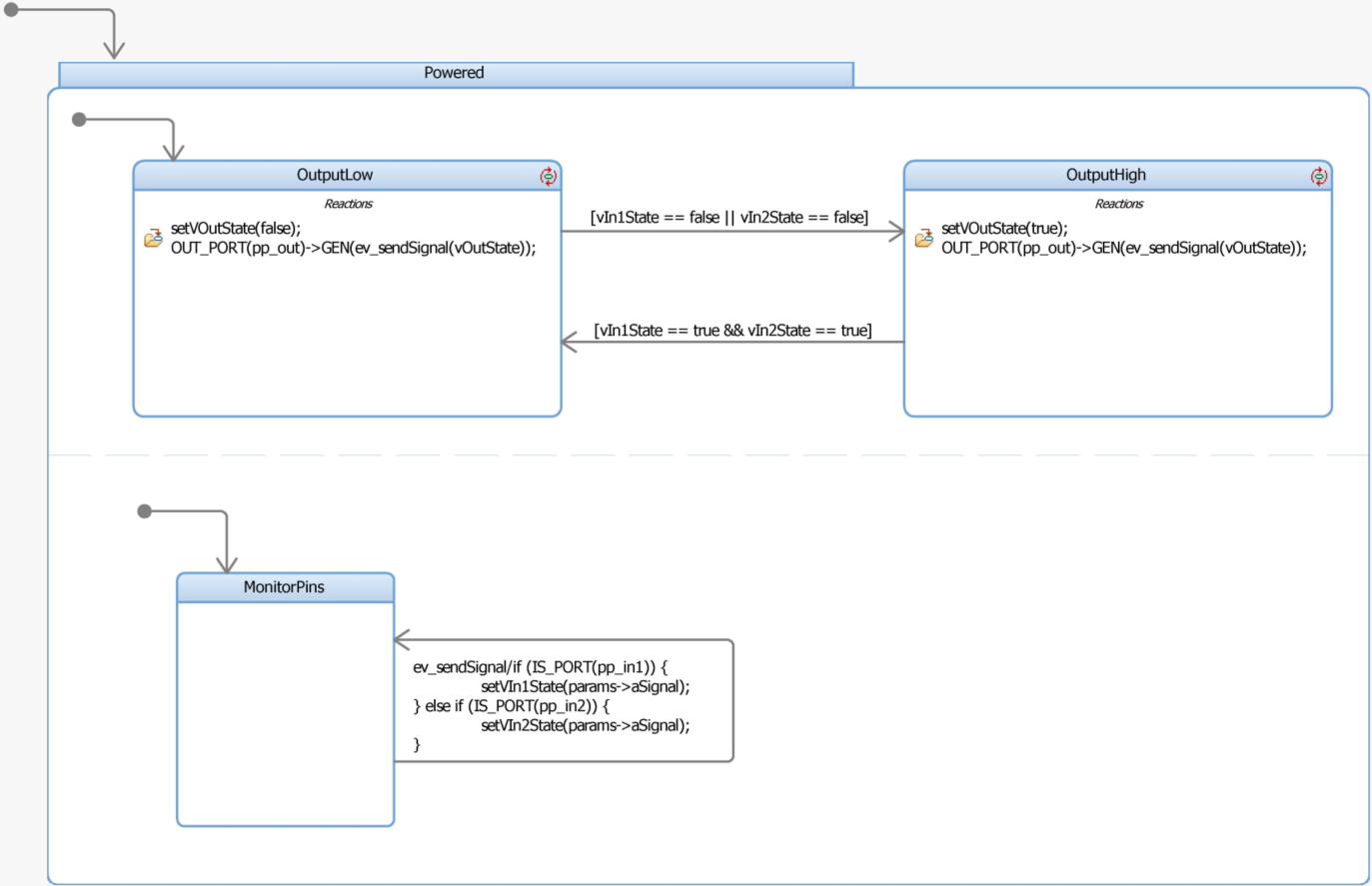}
    \caption{Excutable stateChart of NAND block}
    \label{fig:NAND}
\end{figure}

The corresponding SysML behavioural realisation of this gate block is shown in Figure~\ref{fig:NAND}, where the NAND gate is implemented as an executable statechart rather than as a purely static symbol. This is important because the statechart does not only describe the intended logic of the gate, but also enables that logic to be executed during model simulation. The outer state, \textit{Powered}, represents the condition in which the gate is active and able to receive, process, and transmit signals.

The upper region determines the current output state of the gate, represented by \textit{OutputLow} and \textit{OutputHigh}, while the lower region monitors incoming signal events on the input ports and updates the internal input-state variables accordingly. Based on these stored input values, guarded transitions in the upper region evaluate the NAND condition and activate the appropriate output state. The selected state then assigns the corresponding Boolean value to the output and transmits it through the output port \textit{pp-out}. In this way, the statechart provides the executable mechanism by which the NAND logical relationship is realised dynamically in the SysML model.

\begin{figure*}[!b]
    \centering
    \includegraphics[width=1\linewidth]{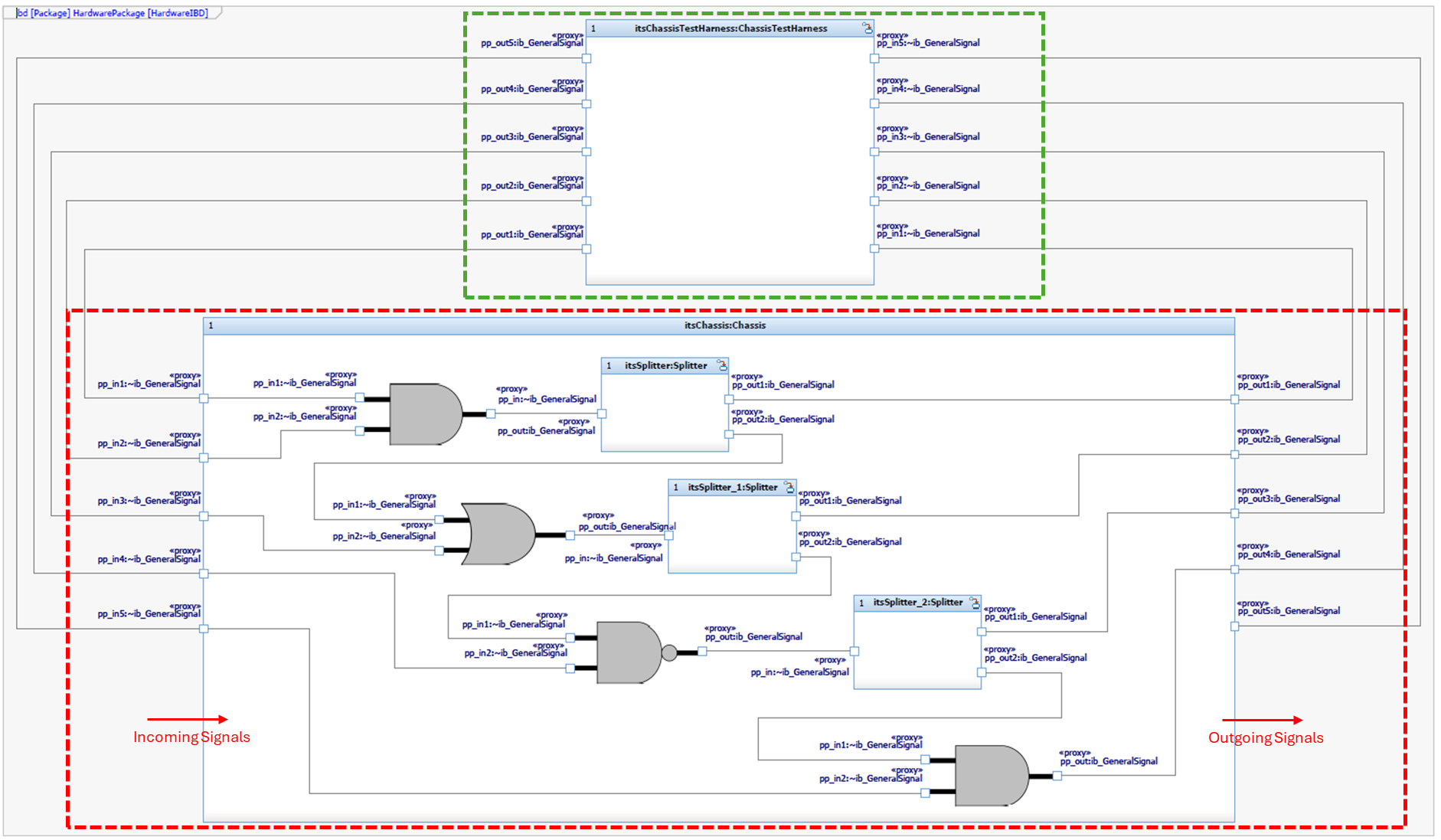}
    \caption{\ac{IBD} diagram of the intended HW prototype with 5 input pins and 5 output pins in red box, and a corresponding test harness in green box}
    \label{fig:Individual}
\end{figure*}

After defining the executable statecharts for the individual logic-gate blocks, it became necessary to introduce additional supporting building blocks before the full hardware model could be assembled. Simple ``one-in, two-out'' splitter blocks were introduced to duplicate a Boolean output signal where one needed to be routed to more than one destination. This was necessary because a signal carried through a single proxy port could not be transmitted simultaneously to multiple receiving ports. These blocks were created as SysML classes so that they could be instantiated as needed within subsystem models.

Once these building blocks had been defined, they were instantiated and interconnected to form the complete hardware model. The intended prototype was specified with five input pins and five output pins. Therefore, an \ac{IBD} was then developed in SysML to define the chassis architecture, its external interfaces, and the internal signal flow between ports before physical implementation. As shown in the red region of Figure~\ref{fig:Individual}, this model provides a simplified but structured representation of the intended physical chassis.

The figure shows the chassis boundary, the external input and output ports, the instantiated logic-gate blocks that perform Boolean operations, and the splitter blocks that distribute signals along multiple internal paths. For clarity, the logic-gate blocks were assigned graphical gate symbols representing their logical function, giving the \ac{IBD} the appearance of a logic arrangement. The connectors between the internal blocks indicate the intended signal paths through the subsystem, showing how external Boolean inputs are processed through the internal arrangement and propagated to the output pins.

\subsubsection{Testing the Hardware Model via the Test Harness}
\label{HWModel}

To control and monitor the pins, a Test Harness was created, as described in Step~\ref{sysMLTest}, and connected to the Chassis \ac{IBD} to provide a stimulus environment for the hardware model by transmitting incoming signals and recording outgoing signals, as shown in the green region of Figure~\ref{fig:Individual}. The Test Harness ports share the same Interface Block as the hardware model, since the same Boolean signal messages are exchanged across all connected ports. Once the overall model had been completed, it could be compiled and executed, enabling the behaviour of the hardware model to be observed and recorded.

This verification step is used to test the hardware model by operating in \ac{MOM}, which allows the operator to toggle the hardware model input pins high or low and monitor the corresponding output pins. In this role, the test harness stimulates the hardware model in place of a more complex system-of-systems model. This means that the same verification approach could be applied to different hardware models, as would typically be required in a real-world operational context.

\begin{figure*}
    \centering
    \includegraphics[width=1\linewidth]{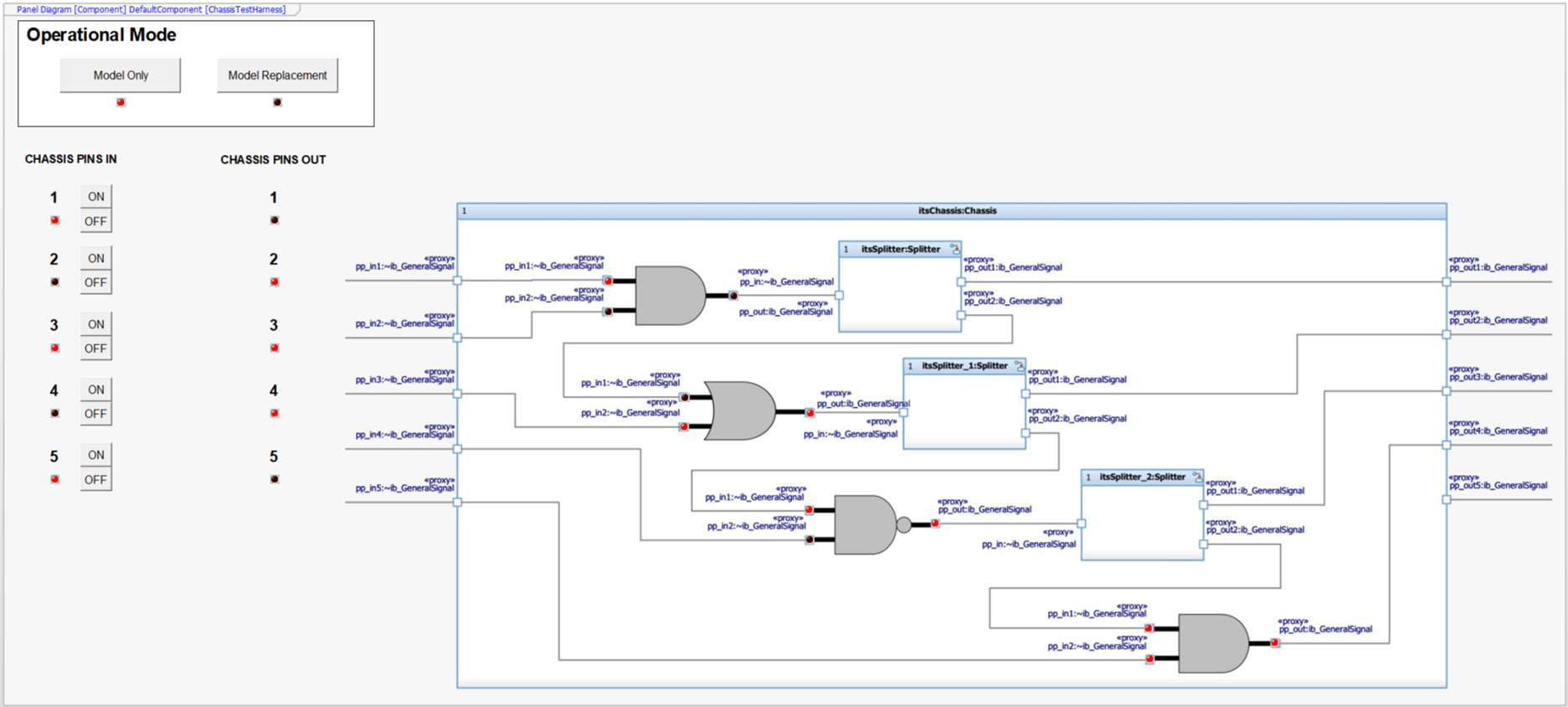}
    \caption{(User interface panel diagram for applying stimuli to and testing the hardware model}
    \label{fig:HwModel}
\end{figure*}

To enable the operator to stimulate the hardware model through the Test Harness, without relying on the full system model, a Panel Diagram was developed, as shown in Figure~\ref{fig:HwModel}. This diagram serves as a user interface within the SysML environment and can be understood through three main parts:

\begin{enumerate}
    \item Operational mode selection, at the top left of Figure~\ref{fig:HwModel}.

 This allows the operator to choose whether the test is directed towards the hardware model or the physical prototype through two defined modes: \ac{MOM} and \ac{MRM}.

    \item CHASSIS PINS IN, on the left-hand side
    
Five input channels are provided, numbered from 1 to 5. Each channel includes ON and OFF controls, enabling the operator to apply manual input stimuli. Selecting ON drives the corresponding input pin high, while selecting OFF drives it low.

    \item CHASSIS PINS OUT \& \ac{IBD}, on the right-hand side

This part illustrates indicator lamps that display the resulting output states after the hardware model processes the selected inputs. In addition, a figure showing the chassis \ac{IBD} previously developed in Figure~\ref{fig:HwModel}. Lamps are placed on the pins of the internal logic gates so that the propagation of signals through the model can be observed directly. This allows the developer to monitor not only the final output states, but also the intermediate internal signal states during testing, which was particularly useful for debugging the logic blocks and verifying their behaviour.
    
\end{enumerate}

\begin{figure}[ht]
    \centering
    \includegraphics[width=1\linewidth]{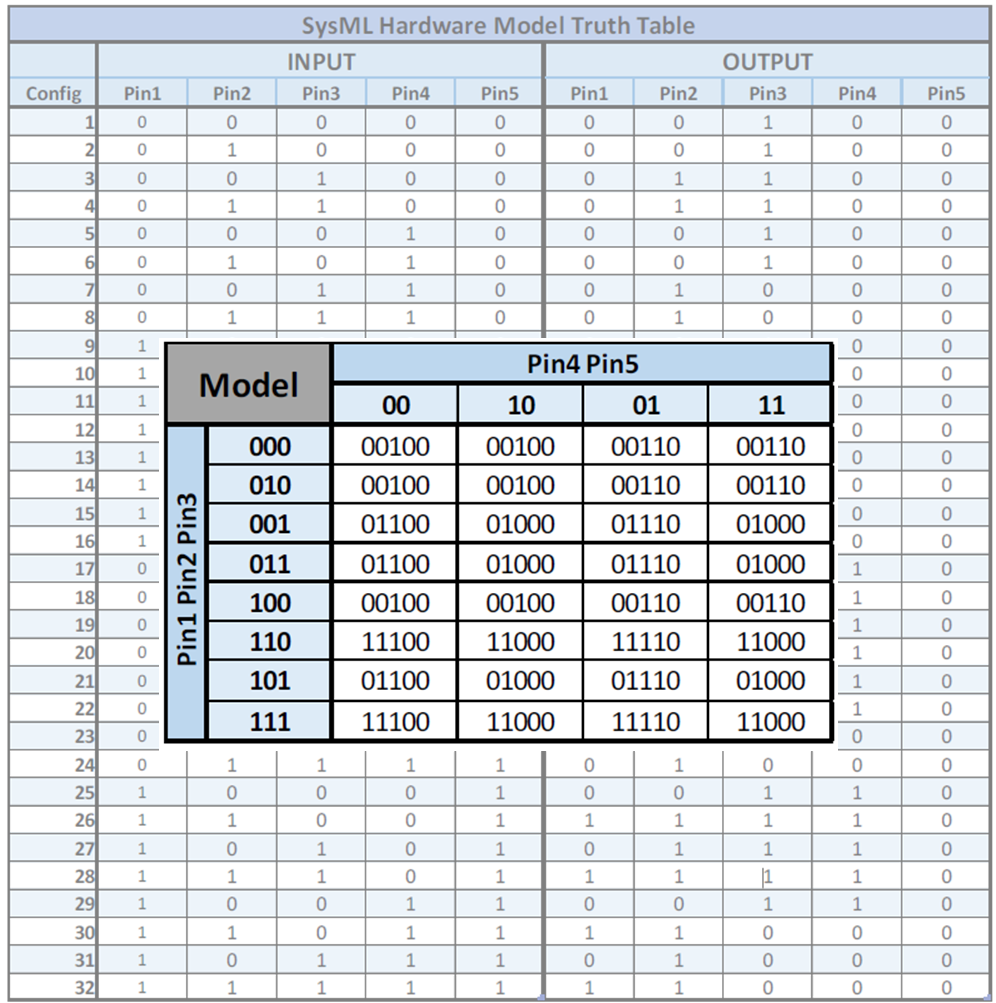}
    \caption{SysML Hardware Model Truth Table and its converting to Karnaugh Map}
    \label{fig:tables}
\end{figure}

This behaviour of the hardware model was assessed through the generation of a truth table. Every possible combination of the five input pins was tested, and the corresponding outputs were simulated and recorded. Since five Boolean inputs give $2^{5} = 32$ possible input combinations, the table contains 32 rows. As this truth table is relatively large and difficult to interpret directly, it was converted into a Karnaugh map in order to reduce its visual complexity and to support easier comparison with the physical hardware at a later stage. This map is shown in Figure~\ref{fig:tables}.

\subsection{Assembling and Testing of a Physical Hardware}

\begin{figure}[ht]
    \centering
    \includegraphics[width=1\linewidth]{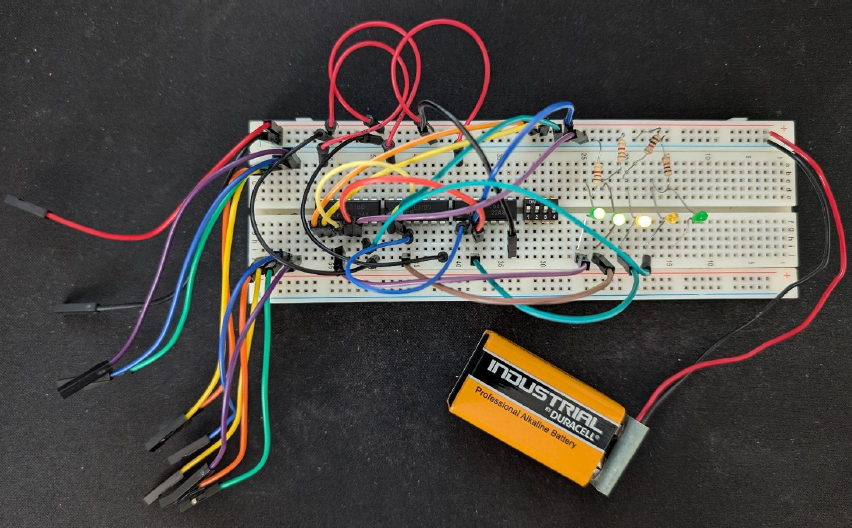}
    \caption{Isolated Hardware Testing Setup}
    \label{fig:isolated}
\end{figure}

The hardware was then assembled, using the SysML model as a direct manufacturing reference, onto a prototyping board. In this physical realisation, the functional logic network defined in the model was implemented using discrete logic integrated circuits and jumper-wire connections. To simplify observation during isolated testing, the five modelled output pins were represented physically by five Light Emitting Diodes (LEDs), each connected in series with a current-limiting resistor. This allowed the output state of the hardware to be read visually, with each illuminated LED indicating an active output condition. This hardware could then be manually tested by applying 9 VDC in all possible configurations across the five input points. The physical setup mid-test can be seen in Figure~\ref{fig:isolated}, with the first 3 of the five LEDs illuminated on the right. The input and output wires on the left are disconnected, and signals are applied directly to the corresponding logic gate pins using the +9VDC rail at the top.

The recorded results from the physical hardware testing were manually checked by the operator to confirm that the assembled physical chassis was functioning as expected, \ref{HWTest}.

\subsection{\ac{SHIA} Servers Development}

As discussed previously, the \ac{SHIA} mechanism is divided into two servers. The first is the \ac{SysML} server, which refers to the C++ implementation associated with the SysML environment and acts as the bridge between the executable SysML model and the external serial communication link through the USB cable. The second is the \ac{SHIA} Hardware Server, which consists of a programmed Raspberry Pi board responsible for interfacing with and controlling the physical hardware.

The initial development of the \ac{SHIA} SysML Server proceeded largely through trial and error, with multiple changes made to both the SysML model and its associated C++ compilation settings. This eventually led to a condition in which the model no longer compiled and the source of the problem could not be readily identified. To address this, development was restarted under stricter configuration control, including versioned model files and a changelog to record major script and configuration changes. This improved traceability, simplified rollback, and made debugging more manageable.

The decision to use serial communication introduced the need for additional hardware to interface the Raspberry Pi with the serial link, discussed in detail in the subsequent steps \ref{HWPrep}. To control the Raspberry Pi hardware interface and enable it to interpret the incoming signals from SysML side, the open-source WiringPi library was used \cite{wiringpi_python}.

The C++ script implementing the \ac{SHIA} SysML Server and the C++ serial server code executed on the Raspberry Pi side are provided separately in the Supplementary Information to support further review and reproducibility.

\subsection{\ac{SHIA} Servers Verification in Isolation}

The purpose of this stage was to verify the \ac{SHIA} mechanism itself, rather than to verify the hardware model or the physical hardware independently. This corresponds to Step~\ref{SHIATest}. To achieve this, each \ac{SHIA} server was first connected to its corresponding counterpart. Accordingly, the \ac{SHIA} Hardware Server was connected to the physical hardware, as shown in Figure~\ref{fig:HW}, while the \ac{SHIA} SysML Server was connected to the Test Harness within the SysML environment. The two servers were then linked through the serial communication connection.

This arrangement was established specifically to verify the operability of the \ac{SHIA} mechanism and its ability to support bidirectional communication between the SysML environment and the physical hardware. Verification was performed by transmitting messages from the SysML side, through the Test Harness, and recording the messages received from the hardware side, in the SysML truth tables.

\subsubsection{\ac{SHIA} Hardware Preparation}
\label{HWPrep}

\begin{figure}[ht]
    \centering
    \includegraphics[width=1\linewidth]{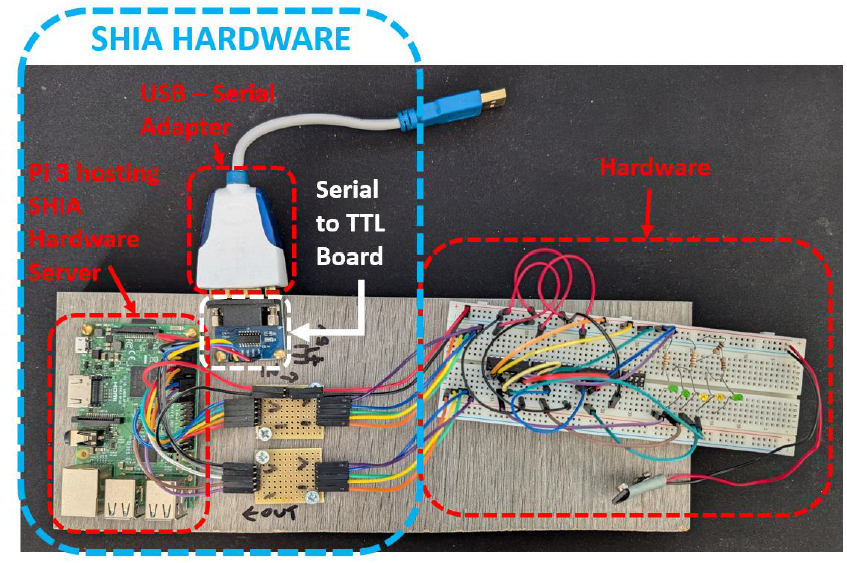}
    \caption{\ac{SHIA} Hardware Server Implementation}
    \label{fig:HW}
\end{figure}

 A USB-to-Serial (male DB-9) cable was procured, together with a MAX3232-based serial (female DB-9) to TTL header converter board, in order to translate the serial signals into low-voltage 3.3--5 V levels suitable for the Raspberry Pi and thereby avoid damage to its GPIO pins. These GPIO pins are configured via a C++ script, auto-executing on the Raspberry Pi, to both interpret incoming signals and enable control of discrete actions on the physical HW board. The resulting \ac{SHIA} hardware arrangement therefore comprised the USB-to-Serial cable, the serial-to-TTL converter board, the programmed Raspberry Pi, and the connecting wires to the prototype hardware, as shown in Figure~\ref{fig:HW}.

\subsubsection{\ac{SHIA} SysML Preparation}

To prepare for \ac{SHIA} server verification, the behavioural workflow was governed directly within the SysML environment. This demonstrates SysML functioning as the \textit{single source of truth}, as the operational logic of the interface was defined and controlled by the SysML model itself. Accordingly, the model served not only as a representation of intended behaviour, but also as the mechanism through which message generation, transmission, and interpretation were directed during execution.


\noindent\textit{\textbf{Message Generation}}\

\begin{figure*}[ht]
    \centering
    \includegraphics[width=1\linewidth]{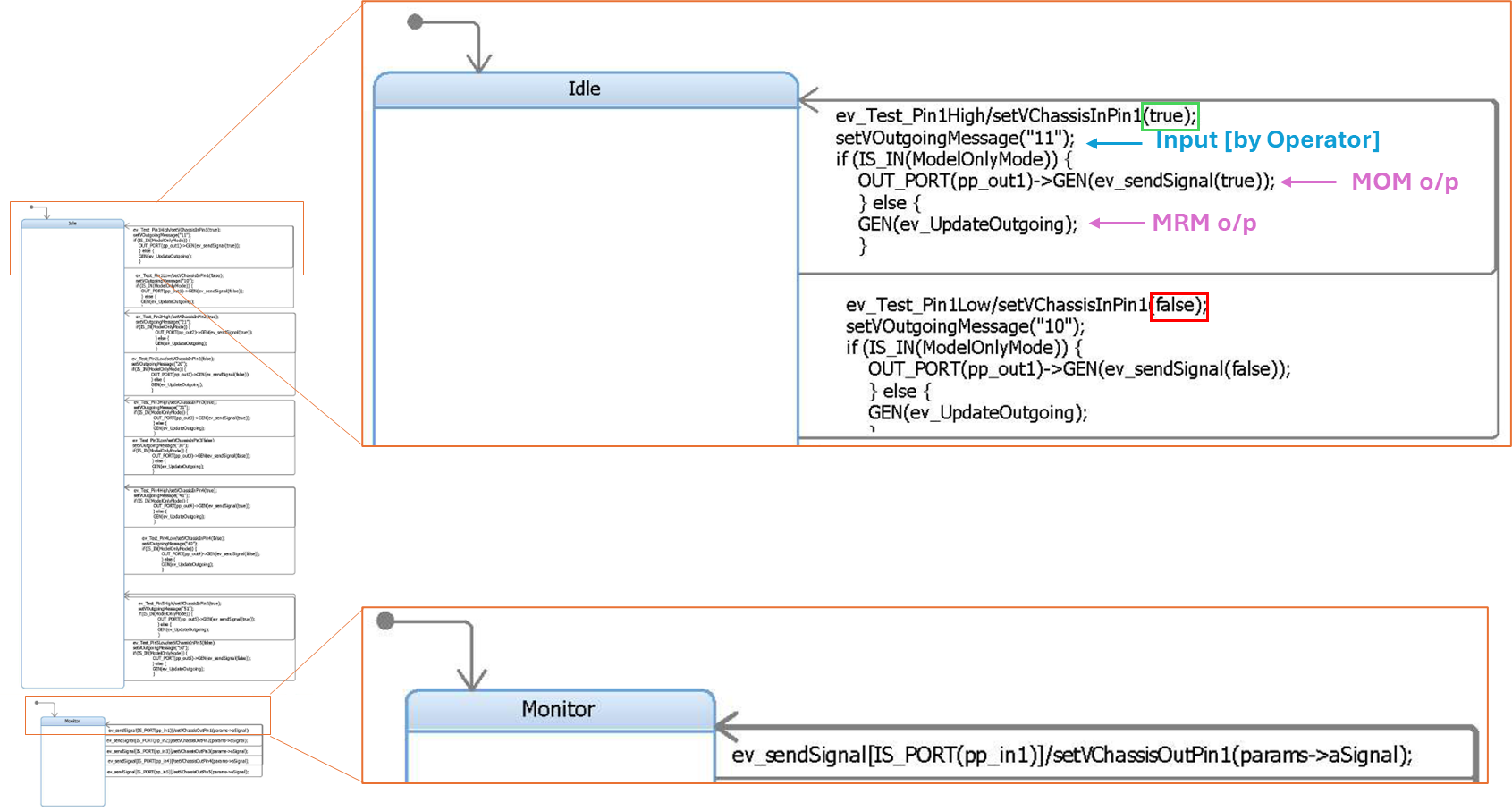}
    \caption{Test Harness behavioural response to operator inputs, showing the concurrent states \texttt{Idle} and \texttt{Monitor}.}
    \label{fig:response}
\end{figure*}

As introduced earlier, the Test Harness was developed not only to verify the hardware model in isolation, but also to demonstrate the operability of \ac{SHIA} in \ac{MRM}. Supporting this dual function, 

 Figure~\ref{fig:response} shows the chart is centred around the \texttt{Idle} state, and \texttt{Monitor} state which acts as a listening state for events generated by the operator-controlled panel diagram. As shown in the zoomed-in view, when an event such as \texttt{ev\_Test\_Pin1High} is received, the corresponding internal attribute, for example \texttt{vChassisInPin1}, is updated to reflect the new state. At the same time, the outgoing message string is assigned the corresponding two-character value, such as \texttt{"11"} to represent pin 1 set high or \texttt{"10"} to represent pin 1 set low.

\begin{figure*}[ht]
    \centering
    \includegraphics[width=1\linewidth]{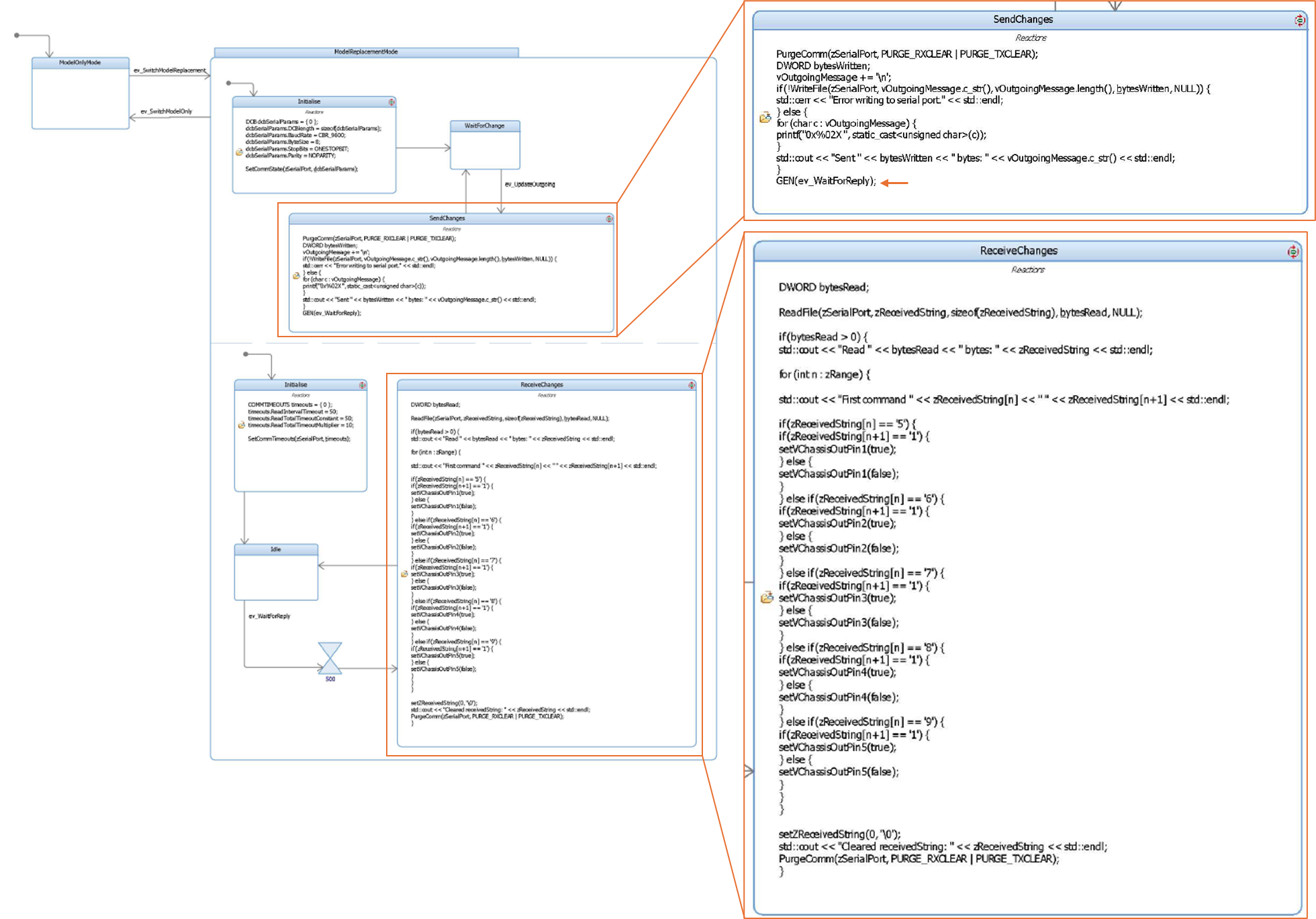}
    \caption{\ac{SHIA} SysML Server Model Interface Transmit State}
    \label{fig:sender}
\end{figure*}

At this point, the chart makes a decision based on the active operating mode. If the Test Harness is running in \ac{MOM}, the change is passed directly to the output port of the Test Harness so that the virtual hardware model is stimulated internally. However, if the system is operating in \ac{MRM}, the event \texttt{ev\_UpdateOutgoing} is generated instead. This event does not itself send the message, but triggers the next stage of the communication process.

The same mechanism is followed for the remaining pins, whereby each event updates its associated input attribute and generates the corresponding message string for transmission through the concurrent statechart.

\noindent\textit{\textbf{Message Transmission}}\

This state region operates in parallel with the listening behaviour of Figure~\ref{fig:response}. Initially, the chart is in \texttt{ModelOnlyMode}. When the operator switches mode through the panel diagram, the chart enters \texttt{ModelReplacementMode}, indicating that communication with the physical hardware is now required.

Figure~\ref{fig:sender} corresponds to \texttt{ModelReplacementMode}, the first state is \texttt{Initialise}, where the serial communication parameters are configured. These include settings such as byte size, stop bits, and parity, mirroring the configuration of the \ac{SHIA} Hardware Server and thereby preparing the serial port for message transmission. Once this initialisation is complete, the state chart moves to \texttt{WaitForChange}, where it remains until the event \texttt{ev\_UpdateOutgoing} is received from the behaviour described in Figure~\ref{fig:response}.

When this event occurs, the chart enters the \texttt{SendChanges} state. In this state, the contents of \texttt{vOutgoingMessage} are transmitted over the serial port using the \texttt{WriteFile} function. After transmission, the event \texttt{ev\_WaitForReply} is generated, as zoomed-in top view. This event marks the end of the transmission step and triggers the receive-side behaviour. Continuing the previous example, if \texttt{"21"} has been prepared to represent input pin 2 being driven high, Figure~\ref{fig:sender} is responsible for sending this command over the serial connection to the hardware server, controlling the physical prototype pins.

\noindent\textit{\textbf{Message Execution}}

Figure~\ref{fig:sender} aims to show how the SysML-side server retrieves the returned hardware state and updates the model accordingly. This figure represents another parallel state region within model replacement mode. At the bottom view of concurrent events, it begins with an \texttt{Initialise} state. This step prepares the serial interface for subsequent reading operations and ensures that the SysML server can wait for incoming hardware data using consistent timing settings. After this initial configuration, the statechart enters the \texttt{Idle} state, which acts as the waiting state until a reply from the hardware side is expected.

Once the event \texttt{ev\_WaitforReply} occurs, the statechart pauses for the specified delay interval of 500~ms before moving to the \texttt{ReceiveChanges} state. The purpose of this delay is to allow sufficient time for the transmitted input change to propagate through the hardware server, the GPIO pins, and the assembled logic board, and for the resulting output state to be returned over the serial interface. After this delay, the chart enters \texttt{ReceiveChanges}, where the \texttt{ReadFile} function is used to read the returned serial message. The received two-character string is then decoded so that the corresponding output-pin attribute in the SysML model can be updated. In this way, the state of the physical hardware is brought back into the SysML environment and made visible through the Test Harness outputs.

The C++ script on the Hardware side contains an infinite loop checking for data availability on the serial port at around 10Hz. Changes on the GPIO pins are detected and actioned via an interrupt handler such that responses to hardware changes are immediate. The same method could be implemented in the SysML Server side to potentially reduce interface latency, but a more methodical approach was intentionally taken for \ac{SHIA} development, so state changes and message transitions could be more easily monitored and verified.

At the end of this section, the decoded hardware results are compared with the corresponding results from the Hardware model \ref{HWModel} and found to be identical. This confirmed consistency between the modelled and physical system behaviour and provided assurance that the hardware implementation was correct.

\subsection{Verification of Integrated System}

\begin{figure*}
    \centering
    \includegraphics[width=1\linewidth]{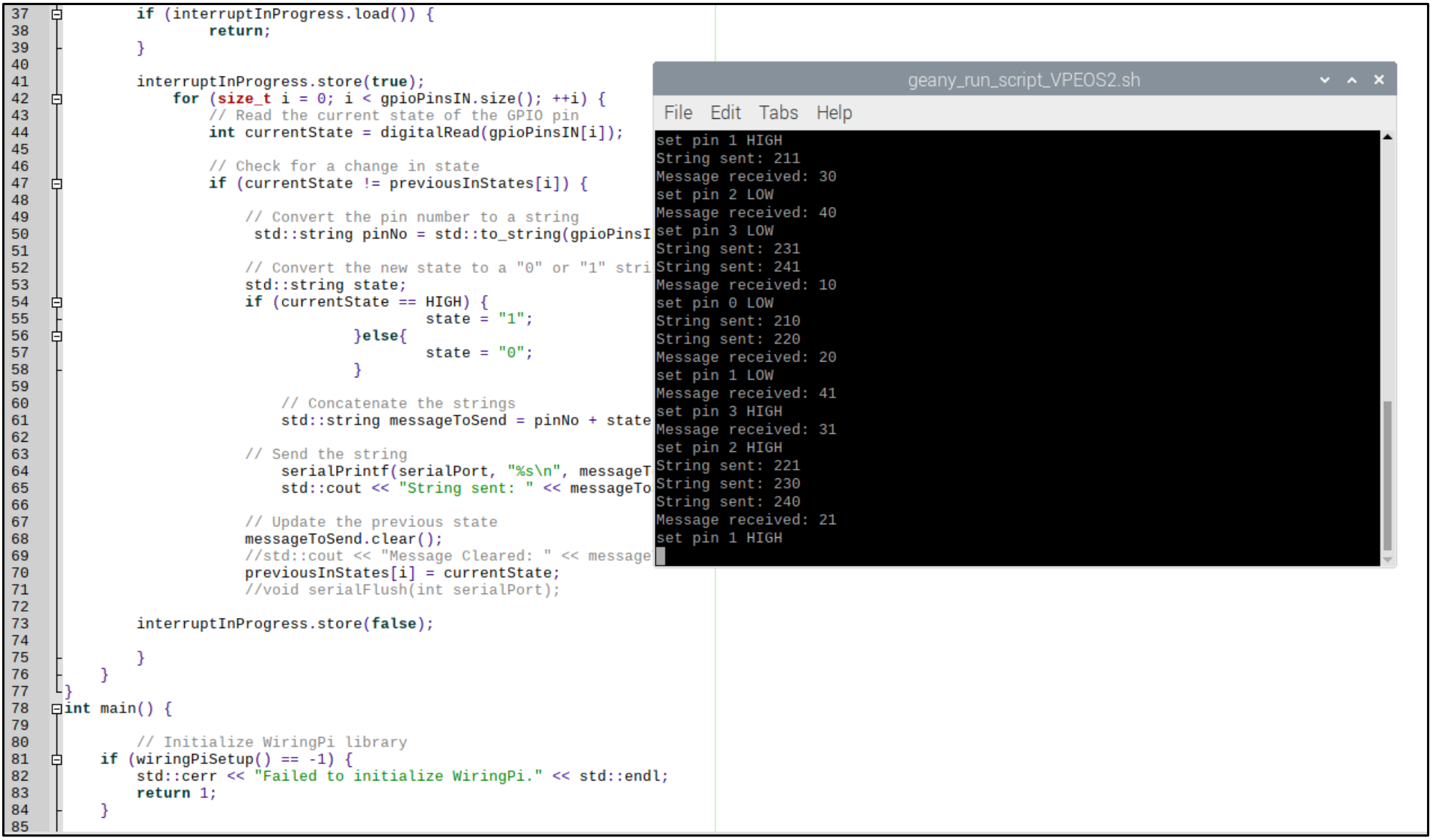}
    \caption{Screenshot of the Hardware Server in operation on the Raspberry Pi}
    \label{fig:serverCode}
\end{figure*}

Because the hardware model, \ac{SHIA} servers, and physical prototype had already undergone rigorous verification refining their composition and codes, integration between the hardware model, \ac{SHIA} SysML server, \ac{SHIA} hardware and the prototype was relatively straightforward and the connection operated consistently.

\begin{figure*}[!b]
    \centering
    \includegraphics[width=1\linewidth]{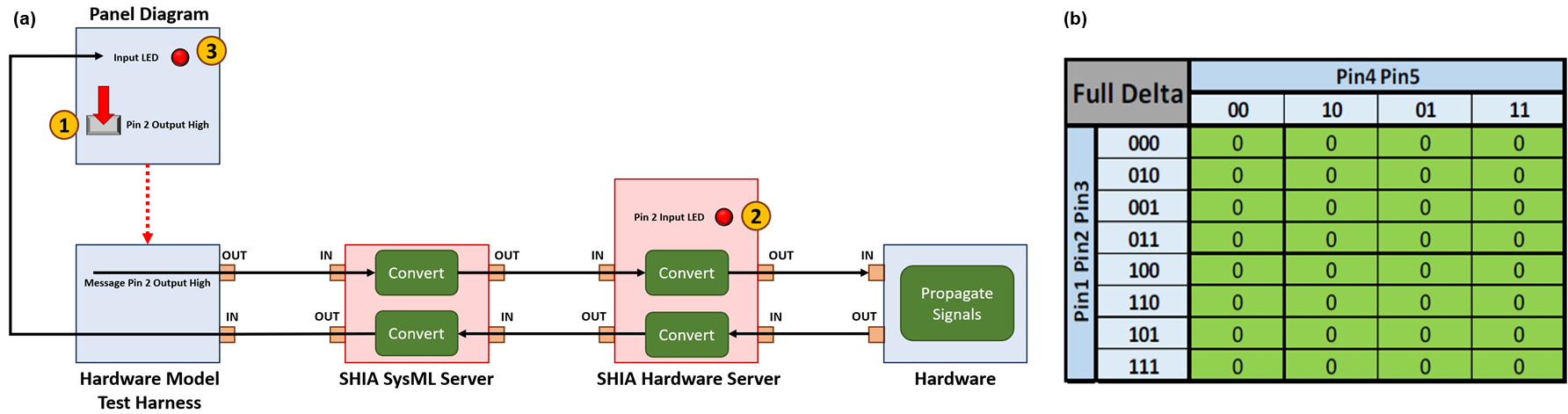}
    \caption{(a) System Verification Method and (b) Karnaugh comparison map showing verification of full \ac{SHIA} system}
    \label{fig:verification}
\end{figure*}

Evidence of successful integration is provided in Figure~\ref{fig:serverCode}, which shows the Hardware Server running on the Raspberry Pi. The shell output demonstrates that command messages were received and executed correctly, for example by setting specific pins HIGH or LOW. It also shows response messages being transmitted after hardware processing, confirming correct communication between the hardware and the SysML server. In this implementation, Raspberry Pi physical GPIO pins 21 to 25 corresponded to SysML input pins 1 to 5.

To verify the functional behaviour of \ac{SHIA}, the same approach was taken as that used to verify the hardware and its model, with the generation of a truth table and Karnaugh map. The difference in this case, however, was that the SysML model was driven via the Panel Diagram, and the hardware output both directly monitored via LED and assessed in terms of the message received back into the SysML model. This is illustrated by Figure~\ref{fig:verification} (a), where the orange numbered circles correspond to the generic steps taken:
\begin{enumerate}
    \item A system model pin is toggled high (in this example Pin 2)
    \item The LED connected to the hardware server GPIO is observed, to ensure that the correct physical pin is triggered
    \item The Panel Diagram system input lights are monitored to ensure that the correct virtual LED lights.
\end{enumerate}

The results of this testing are presented via the Karnaugh comparison map in Figure~\ref{fig:verification}(b). The figure shows the zero result from subtracting one map from the other, proving that when the SysML Test Harness was stimulated by the operator, the message conversions performed by both the SysML server and the Hardware Server were correctly executed in both communication directions, resulting in the correct lamp indication being activated in the SysML Test Harness. This verification therefore demonstrated that \ac{SHIA} operated correctly by transmitting pin-state changes from the simulated SysML model to the physical hardware, and by updating the SysML model in response to changes detected at the physical pins in real time.

These results provide a practical demonstration of SysML operating as the single source of truth, since both virtual stimulation and physical hardware interaction were coordinated from the same authoritative model. They also show that \ac{SHIA} functions as an operable direct interface between SysML and hardware, without requiring model transformation or reliance on an external simulation environment.

\section{Results \& Discussion}
\label{discussion}

\begin{figure*}[ht]
    \centering
    \includegraphics[width=1\linewidth]{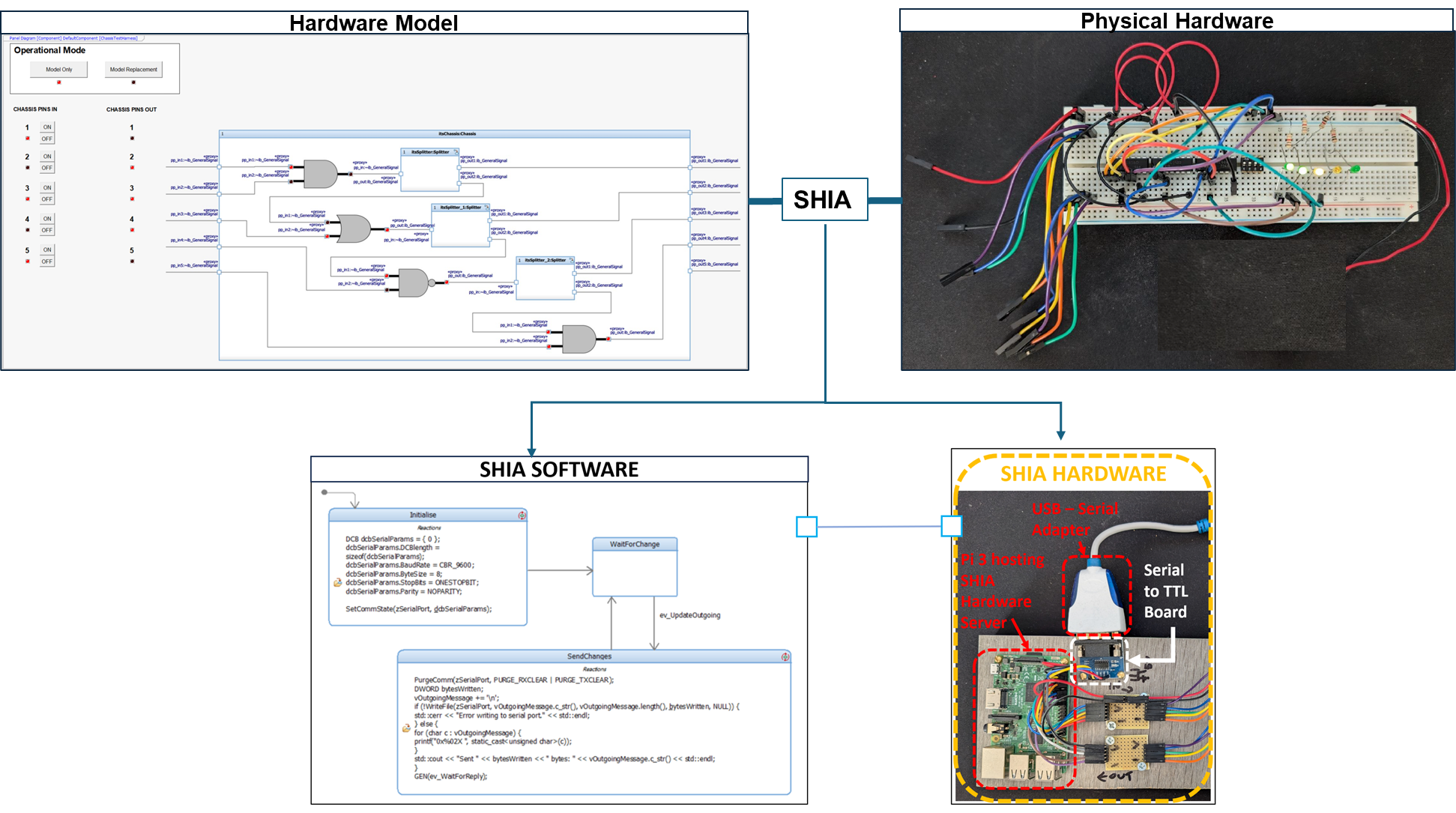}
    \caption{\ac{SHIA} mechanism as a direct real time interface between virtual and physical counterparts}
    \label{fig:SHIADirect}
\end{figure*}

The results demonstrate that \ac{SHIA} provides a feasible mechanism for enabling direct, real-time communication between a SysML model and physical hardware. By exploiting SysML’s native C++ capability to implement a server controlled through a SysML state charts, the study shows that SysML can operate as more than a static modelling environment. Instead, it can take an active role in system execution and interaction with hardware. This is significant because it begins to address a longstanding limitation in \ac{MBSE}, where system models often remain detached from the physical systems they are intended to represent. The \ac{SHIA} approach, therefore, offers a solution to the direct model-to-hardware interaction. This solution promises a coherent way of preserving alignment between design intent and implemented behaviour as witnessed in Figure~\ref{fig:SHIADirect}.

SysML diagrams were used not only to develop the virtual hardware model, but also to capture the behavioural realisation of the \ac{SHIA} mechanism, the Test Harness, user interaction through panel diagrams, and the recording of outputs through truth tables. This integrated use of SysML meant that the same model supported both system development and verification activities.

The findings also highlight the broader role that a behavioural SysML model can play in \ac{VV}. In this study, the model did not serve solely as a design description, but also as an executable behavioural reference against which hardware behaviour could be assessed. Rather than relying exclusively on extensive textual requirement sets as the basis for verification, the model itself provided an explicit and testable representation of expected behaviour. This indicates that behavioural models may contribute more directly to assurance activities than is often assumed in traditional \ac{MBSE} workflows.

This broader use of SysML further reinforced its role as the single source of truth in a digital thread spanning from the development to the verification processes, because it was used without requiring model transformation or parallel intermediary artefacts. As a result, the approach reduced the risk of inconsistency between representations, minimised synchronisation effort, and improved traceability across the modelling, execution, and verification stages. In this way, SysML served not only as a modelling environment, but also as an executable environment through which the evolving hardware could be stimulated, observed, and verified.

\begin{figure}[ht]
    \centering
    \includegraphics[width=0.9\linewidth]{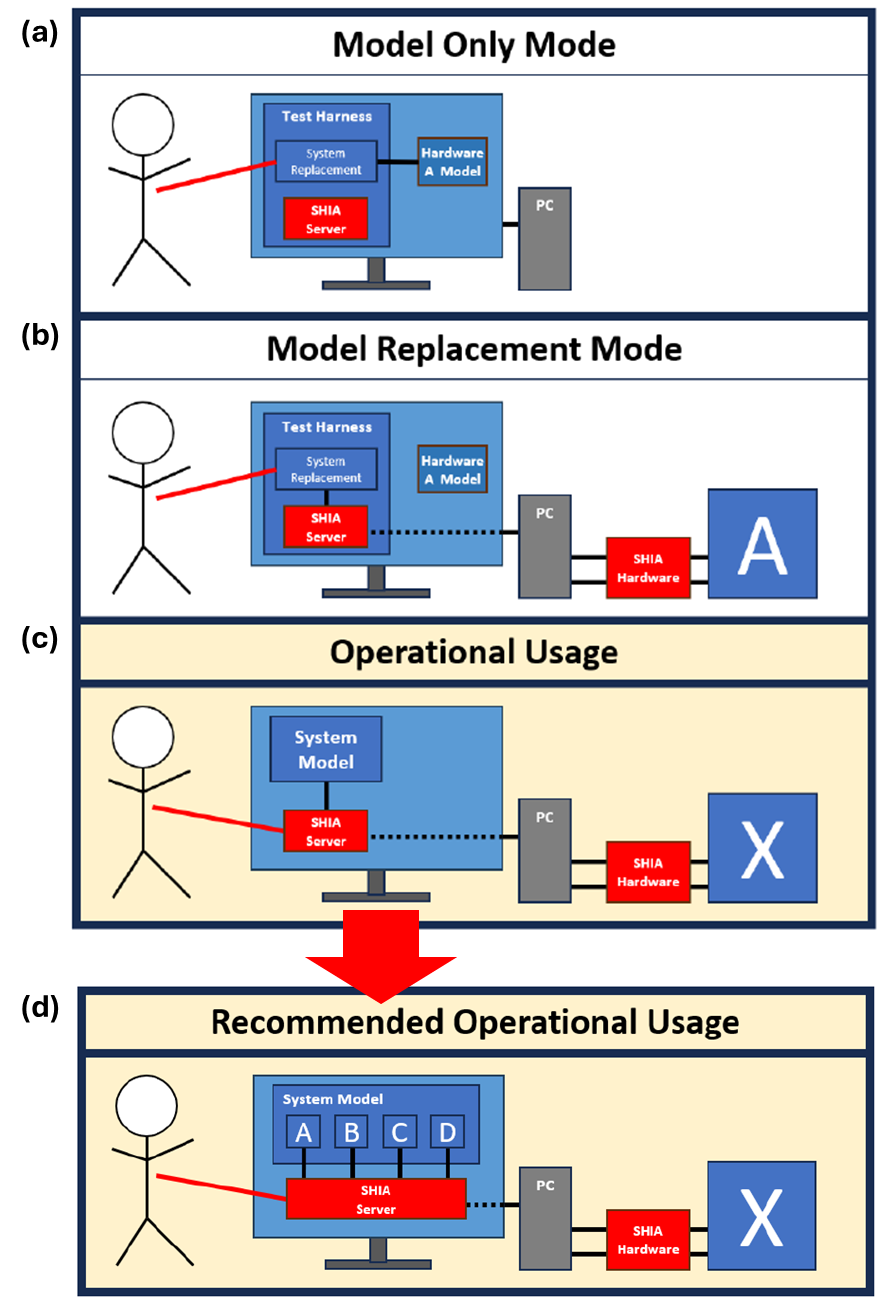}
    \caption{Test Harness Context and Comparison to Forecast Operational Usage}
    \label{fig:testHarness}
\end{figure}

The case study was developed as a proof of concept, following the proposed approach and achieving the expected outcomes required from \ac{SHIA}. The implemented interface was designed specifically for a single hardware model. However, in a full system, the \ac{SHIA} Server would be required to exchange a much larger number of messages across multiple subsystems, covering a broader range of interactions. Accordingly, it was considered more suitable for the \ac{SHIA} server interface to be defined directly at the subsystem level, rather than through a single shared interface. Although a single shared interface may initially appear simpler, it is likely to become difficult to manage if required to contain a large number of message definitions. The resulting recommendation for this interface arrangement is illustrated in Figure~\ref{fig:testHarness}.

For more complex systems, where specialised simulation tools are needed to represent detailed physical behaviour and parameter interactions, SysML may still be preserved as the single source of truth. In such cases, the SysML model need not perform all low-level physical simulation itself, but can remain the authoritative reference model through which system structure, behaviour, requirements, and verification relationships are coordinated. Simulation outputs can then be brought back into the SysML environment, for example through truth tables or comparable result-capturing artefacts, using APIs or implementation scripts in languages such as C++, Java, and Python. This allows SysML to retain governance of the system model while interoperating with domain-specific simulation tools.

A related implication concerns the growing interest in integrating Large Language Models (LLMs) into \ac{MBSE} workflows \cite{Zhang2026MBSECoPilot}. Current efforts in this direction face the recurring challenge that LLMs, when operating on purely textual or loosely structured artefacts, can produce outputs that are difficult to verify against the underlying system \cite{Li2025LLMSysMLv2}. By keeping SysML active as an executable and semantically grounded reference, the \ac{SHIA} approach contributes to the substrate that such tools require for reliable participation in verification activities, although exploring this integration is beyond the scope of the present work.

Taken together, the results suggest that \ac{SHIA} has value not only as a technical interface concept but also as a contribution to the wider discussion on how SysML models can remain operationally relevant throughout the system lifecycle. The study indicates that direct SysML–hardware integration can strengthen continuity between modelling, implementation, and verification, while also reducing fragmentation between engineering artefacts. However, its current maturity remains at the proof-of-concept stage, and its ultimate usefulness will depend on whether the same principles can be extended to larger and more representative systems.



\section{Conclusions}
\label{Conclusion}

This paper has addressed a key limitation in current digital-engineering and \ac{HiL} workflows: the tendency for SysML models to become detached from active verification once domain-specific executable models or hardware test environments are introduced. In such workflows, SysML often captures requirements, architecture, behaviour, interfaces, and verification intent, while simulation and test environments provide the executable capability required for hardware-oriented assessment. Although this arrangement supports a digital-thread view of connected engineering artefacts, the connection between the system model and verification evidence can remain dependent on transformation chains, tool integrations, or manual traceability. This creates a risk that the SysML model no longer reflects the verified hardware behaviour as the system evolves.

To address this gap, the paper proposed the \ac{SHIA}, a direct SysML--hardware integration approach that enables the SysML model to remain active within the verification loop. Rather than treating SysML as a static design artefact that only informs downstream tools, \ac{SHIA} allows the model to act as an operational verification layer and a central reference, or single source of truth, against which physical hardware behaviour can be stimulated, observed, and assessed. In this sense, the proposed approach shortens the conventional digital-thread chain between architecture definition and hardware verification by creating a direct, real-time connection between the system model and its physical counterpart.

The proof-of-concept logic-gate case study demonstrated the feasibility of this approach. A SysML hardware model, test harness, SysML-side server, hardware-side server, and physical prototype were developed and verified through staged testing. The integrated results confirmed correct bidirectional message exchange between the executable SysML model and the physical hardware. The comparison between the SysML-generated and hardware-generated Karnaugh maps showed zero discrepancy, demonstrating behavioural equivalence between the digital and physical counterparts in the implemented case study.

The findings show that SysML can support more than descriptive modelling when combined with an appropriate interface mechanism. It can also provide an executable and traceable environment for hardware interaction and verification. This strengthens continuity between system design, implementation, and verification, reduces the likelihood of model--hardware divergence, and supports the co-evolution of system models and physical subsystems. The work therefore contributes to digital engineering by demonstrating a practical route for keeping system architecture and hardware verification directly connected within the digital thread, establishing a foundation for model-governed, digital-twin-like verification in which SysML remains actively connected to the physical system throughout development and testing.


\section*{Acknowledgement}

The authors gratefully acknowledge Melanie King, Michael Henshaw, and Paul Heath for their helpful comments and discussions during the development of this work.

\section*{CRediT authorship contribution statement}

\textbf{Charles Lewis:} Conceptualisation, Methodology, Software, Investigation, Writing -- Original Draft. \textbf{Amal Elsokary:} Visualisation, Validation, Writing -- Original Draft, Writing -- Review \& Editing. \textbf{Siyuan Ji:} Conceptualisation, Supervision, Writing -- Review \& Editing.

\section*{Declaration of generative AI}

During the preparation of this work, the authors used generative AI tools only to support language refinement, grammar improvement, and clarity of expression. The tools were not used to generate the research ideas, methodology, implementation, results, analysis, or conclusions. All intellectual content, technical decisions, implementation, interpretation of results, and final manuscript content were developed, verified, and approved by the authors. The authors take full responsibility for the content of the published article.

\section*{Declaration of competing interest}

The authors declare that they have no known competing financial interests or personal relationships that could have appeared to influence the work reported in this paper.

\section*{Data availability}


The data, SysML models, and C++ source code supporting the findings of this study are openly available in the GitHub repository: \url{https://github.com/AmalElsokary/HiL}.

\bibliographystyle{elsarticle-num} 
\bibliography{References}

@article{sprock2026bridging,
  title={Bridging Model-Based Systems Engineering, Digital Twins, and Cyber-Physical Manufacturing Systems: A Foundational Framework for Operational Excellence},
  author={Sprock, Oscar Silva and others},
  journal={International Multidisciplinar Journal of Emerging Technologies and Applications},
  volume={1},
  number={1},
  pages={61--96},
  year={2026}
}

@incollection{pessoa2022model,
  title={Model-based digital threads for socio-technical systems},
  author={Pessoa, Marcus Vinicius Pereira and Pires, Lu{\'\i}s Ferreira and Moreira, Jo{\~a}o Luiz Rebelo and Wu, Chunlong},
  booktitle={Machine learning for smart environments/cities: An IoT approach},
  pages={27--52},
  year={2022},
  publisher={Springer}
}

@article{fieber2014assessing,
  title={Assessing usability of model driven development in industrial projects},
  author={Fieber, Florian and Regnat, Nikolaus and Rumpe, Bernhard},
  journal={arXiv preprint arXiv:1409.6588},
  year={2014}
}

@article{nolan2008model,
  title={Model driven systems development with rational products},
  author={Nolan, Brian and Brown, Barclay and Balmelli, Laurent and Bohn, Tim and Wahli, Ueli},
  journal={IBM Redbooks},
  year={2008}
}

@article{huang2020towards,
  title={Towards digital engineering: the advent of digital systems engineering},
  author={Huang, Jingwei and Gheorghe, Adrian and Handley, Holly and Pazos, Pilar and Pinto, Ariel and Kovacic, Samuel and Collins, Andrew and Keating, Charles and Sousa-Poza, Andres and Rabadi, Ghaith and others},
  journal={International Journal of System of Systems Engineering},
  volume={10},
  number={3},
  pages={234--261},
  year={2020},
  publisher={Inderscience Publishers (IEL)}
}

@misc{speedgoat2025solutions,
  author       = {{Speedgoat}},
  title        = {Solutions},
  year         = {2025},
  howpublished = {\url{https://www.speedgoat.com/solutions}},
  note         = {Accessed: 14 Oct 2025}
}

@article{nigischer2021multi,
  title={Multi-domain simulation utilizing SysML: state of the art and future perspectives},
  author={Nigischer, Christian and Bougain, S{\'e}bastien and Riegler, Rainer and Stanek, Heinz Peter and Grafinger, Manfred},
  journal={Procedia CIRP},
  volume={100},
  pages={319--324},
  year={2021},
  publisher={Elsevier}
}

@article{chabibi2018towards,
  title={Towards a Model Integration from SysML to MATLAB/Simulink.},
  author={Chabibi, Bassim and Anwar, Adil and Nassar, Mahmoud},
  journal={J. Softw.},
  volume={13},
  number={12},
  pages={630--645},
  year={2018}
}

@article{barbau2019translator,
  title={Translator from extended SysML to physical interaction and signal flow simulation platforms},
  author={Barbau, Raphael and Bock, Conrad and Dadfarnia, Mehdi},
  journal={Journal of Research of the National Institute of Standards and Technology},
  volume={124},
  pages={1},
  year={2019}
}

@techreport{omg2021sysphs,
  author       = {{Object Management Group (OMG)}},
  title        = {SysML Extension for Physical Interaction and Signal Flow Simulation (SysPhS), Version 1.1},
  institution  = {Object Management Group},
  year         = {2021},
  url          = {https://www.omg.org/spec/SysPhS/1.1/},
  note         = {Accessed: 14 Oct 2025}
}

@misc{Jankevicius2020SysPhS,
  author       = {Nerijus Jankevicius},
  title        = {{OMG SysPhS: Integrating SysML, Simulink, Modelica and FMI}},
  year         = {2020},
  howpublished = {Presentation, INCOSE International Workshop, Torrance, CA, 27 January 2020},
  organization = {CATIA | No Magic},
  url          = {https://omgwiki.org/MBSE/lib/exe/fetch.php?media=mbse:smswg:smswg_20:omg_sysphs_-_sysml_simulink_modelica_fmi_incose.pdf},
  note         = {Accessed: 2026-03-23}
}

@techreport{fmi2023spec,
  author       = {{Modelica Association}},
  title        = {Functional Mock-up Interface Specification, Version 3.0},
  institution  = {Modelica Association Project FMI},
  year         = {2023},
  url          = {https://fmi-standard.org/docs/3.0/},
  note         = {Accessed: 14 Oct 2025}
}

@inproceedings{wang2013hybridsim,
  title={Hybridsim: A modeling and co-simulation toolchain for cyber-physical systems},
  author={Wang, Baobing and Baras, John S},
  booktitle={2013 IEEE/ACM 17th International Symposium on Distributed Simulation and Real Time Applications},
  pages={33--40},
  year={2013},
  organization={IEEE}
}

@article{kalawsky2013bridging,
  title={Bridging the gaps in a model-based system engineering workflow by encompassing hardware-in-the-loop simulation},
  author={Kalawsky, Roy S and O'Brien, John and Chong, Seng and Wong, ChiBiu and Jia, Haibo and Pan, Hongtao and Moore, Philip R},
  journal={IEEE systems Journal},
  volume={7},
  number={4},
  pages={593--605},
  year={2013},
  publisher={IEEE}
}

@inproceedings{rossa2024epics,
  title={EPICS Integration for Rapid Control Prototyping Hardware from Speedgoat},
  author={Rossa, Lutz and Brendike, Maxim and others},
  booktitle={19th International Conference on Accelerator and Large Experimental Physics Control Systems (ICALEPCS'23), Cape Town, South Africa, 09-13 October 2023},
  pages={1317--1321},
  year={2024},
  organization={JACOW Publishing, Geneva, Switzerland}
}

@inproceedings{godart2017generating,
  title={Generating real-time robotics control software from SysML},
  author={Godart, Peter and Gross, Johannes and Mukherjee, Rudranarayan and Ubellacker, Wyatt},
  booktitle={2017 IEEE Aerospace Conference},
  pages={1--11},
  year={2017},
  organization={IEEE}
}

@inproceedings{wang2019model,
  title={A Model-Based V\&V Test Strategy Based on Emerging System Modeling Techniques},
  author={Wang, Gan and Pavalkis, Saulius},
  booktitle={INCOSE International Symposium},
  volume={29},
  number={1},
  pages={771--787},
  year={2019},
  organization={Wiley Online Library}
}

@inproceedings{hoyos2011hiles2,
  title={HiLeS2: model driven embedded system virtual prototype generation},
  author={Hoyos, Horacio and Casallas, Rubby and Jim{\'e}nez, Fernando and Correal, Dar{\'\i}o},
  booktitle={Proceedings of the 2011 Symposium on Theory of Modeling \& Simulation: DEVS Integrative M\&S Symposium},
  pages={75--82},
  year={2011}
}

@inproceedings{gutierrez2014hardware,
  title={Hardware-in-the-loop based sysml for model and control design of interleaved boost converters},
  author={Gutierrez, A and Chamorro, HR and Jimenez, Jose Fernando},
  booktitle={2014 IEEE 15th Workshop on Control and Modeling for Power Electronics (COMPEL)},
  pages={1--6},
  year={2014},
  organization={IEEE}
}

@inproceedings{gutierrez2015hardware,
  title={Hardware-in-the-loop simulation of PV systems in micro-grids using SysML models},
  author={Gutierrez, A and Chamorro, HR and Jimenez, JF and Villa, Luiz Fernando Lavado and Alonso, Corinne},
  booktitle={2015 IEEE 16th Workshop on Control and Modeling for Power Electronics (COMPEL)},
  pages={1--5},
  year={2015},
  organization={IEEE}
}

@inproceedings{gudemann2010sysml,
  title={SysML in digital engineering},
  author={G{\"u}demann, Matthias and Kegel, Stefan and Ortmeier, Frank and Poenicke, Olaf and Richter, Klaus},
  booktitle={Proceedings of the First International Workshop on Digital Engineering},
  pages={1--8},
  year={2010}
}

@article{bajaj2022systems,
  title={Systems modeling language (SysML v2) support for digital engineering},
  author={Bajaj, Manas and Friedenthal, Sanford and Seidewitz, Ed},
  journal={Insight},
  volume={25},
  number={1},
  pages={19--24},
  year={2022},
  publisher={Wiley Online Library}
}

@article{jones2020characterising,
  title={Characterising the Digital Twin: A systematic literature review},
  author={Jones, David and Snider, Chris and Nassehi, Aydin and Yon, Jason and Hicks, Ben},
  journal={CIRP journal of manufacturing science and technology},
  volume={29},
  pages={36--52},
  year={2020},
  publisher={Elsevier}
}

@article{grieves2022digital,
  title={Digital twin: manufacturing excellence through virtual factory replication. 2014},
  author={Grieves, Michael},
  journal={White Paper},
  year={2022}
}

@inproceedings{pierce2023orion,
  title={Orion SysML model, digital twin, and lessons learned for artemis I},
  author={Pierce, Gregory J and Heeren, Joshua D and Hill, Terry R},
  booktitle={INCOSE International Symposium},
  volume={33},
  number={1},
  pages={290--304},
  year={2023},
  organization={Wiley Online Library}
}

@article{carroll2016systematic,
  title={Systematic literature review: How is model-based systems engineering justified?},
  author={Carroll, Edward Ralph and Malins, Robert Joseph},
  year={2016},
  publisher={Sandia National Laboratories (SNL-NM), Albuquerque, NM (United States~…},
  journal={INCOSE Website}
}

@inproceedings{harvey2012document,
  title={Document the model, don’t model the document},
  author={Harvey, David and Waite, M and Logan, P and Liddy, Tommie},
  booktitle={Proc. Syst. Eng./Test Eval. Conf. 6th Asia Pac. Conf. Syst. Eng},
  year={2012}
}

@article{cederbladh2024early,
  title={Early validation and verification of system behaviour in model-based systems engineering: A systematic literature review},
  author={Cederbladh, Johan and Cicchetti, Antonio and Suryadevara, Jagadish},
  journal={ACM Transactions on Software Engineering and Methodology},
  volume={33},
  number={3},
  pages={1--67},
  year={2024},
  publisher={ACM New York, NY, USA}
}

@ARTICLE{ISOStandard,
  author={ISO/IEC/IEEE},
  journal={ISO/IEC/IEEE 15288:2023(E)}, 
  title={ISO/IEC/IEEE International Standard - Systems and software engineering--System life cycle processes}, 
  year={2023},
  volume={},
  number={},
  pages={1-128},
  doi={10.1109/IEEESTD.2023.10123367}}

@inproceedings{miller2022future,
  title={The Future of Systems Engineering: Realizing the Systems Engineering Vision 2035.},
  author={Miller, William D and others},
  booktitle={TE},
  pages={739--747},
  year={2022}
}

@inproceedings{akundi2022perceptions,
  title={Perceptions and the extent of Model-Based Systems Engineering (MBSE) use--An industry survey},
  author={Akundi, Aditya and Ankobiah, Wilma and Mondragon, Oscar and Luna, Sergio},
  booktitle={2022 IEEE International Systems Conference (SysCon)},
  pages={1--7},
  year={2022},
  organization={IEEE}
}

@inproceedings{hause2006sysml,
  title={The SysML modelling language},
  author={Hause, Matthew and others},
  booktitle={Fifteenth European systems engineering conference},
  volume={9},
  pages={1--12},
  year={2006}
}

@inproceedings{fosse2013systems,
  title={Systems engineering interfaces: A model based approach},
  author={Fosse, Elyse and Delp, Christopher L},
  booktitle={2013 IEEE Aerospace Conference},
  pages={1--8},
  year={2013},
  organization={IEEE}
}

@misc{wiringpi_python,
  author       = {{WiringPi}},
  title        = {{WiringPi-Python}},
  year         = {2023},
  howpublished = {\url{https://github.com/WiringPi/WiringPi-Python}},
  note         = {GitHub repository, archived on 31 October 2023, accessed 18 March 2026}
}

@online{DassaultSysMLPlugin,
  author       = {{Dassault Syst\`emes}},
  title        = {No Magic SysML Plugin},
  year         = {2026},
  url          = {https://www.3ds.com/products/catia/no-magic/sysml-plugin},
  note         = {Accessed: 2026-03-23}
}

@online{IBMRhapsody,
  author       = {{IBM}},
  title        = {IBM Engineering Rhapsody},
  year         = {2026},
  url          = {https://www.ibm.com/products/engineering-rhapsody},
  note         = {Accessed: 2026-03-23}
}

@article{de2022taxonomy,
  title={A taxonomy of MBSE approaches by languages, tools and methods},
  author={De Saqui-Sannes, Pierre and Vingerhoeds, Rob A and Garion, Christophe and Thirioux, Xavier},
  journal={IEEE Access},
  volume={10},
  pages={120936--120950},
  year={2022},
  publisher={IEEE}
}

@inproceedings{yeiser2024exploring,
  title={Exploring the executable SYSML capabilities to integrate and operate hardware in the loop},
  author={Yeiser, Alexander and Pavalkis, Saulius and Yakimenko, Oleg},
  booktitle={INCOSE International Symposium},
  volume={34},
  number={1},
  pages={2489--2508},
  year={2024},
  organization={Wiley Online Library}
}

@inproceedings{helle2024hardware,
  title={Hardware-in-the-Loop with SysML and Cameo Systems Modeler},
  author={Helle, Philipp and Schramm, Gerrit},
  booktitle={INCOSE International Symposium},
  volume={34},
  number={1},
  pages={1807--1819},
  year={2024},
  organization={Wiley Online Library}
}

@article{kiesbye2019hardware,
  title={Hardware-in-the-loop and software-in-the-loop testing of the move-ii cubesat},
  author={Kiesbye, Jonis and Messmann, David and Preisinger, Maximilian and Reina, Gonzalo and Nagy, Daniel and Schummer, Florian and Mostad, Martin and Kale, Tejas and Langer, Martin},
  journal={Aerospace},
  volume={6},
  number={12},
  pages={130},
  year={2019},
  publisher={MDPI}
}

@inproceedings{gomez2010embedded,
  title={Embedded systems requirements verification using HiLeS designer},
  author={Gomez, CE and Pascal, JC and Esteban, P and Deleris, Y and Devatine, JR},
  booktitle={ERTS2 2010, Embedded Real Time Software \& Systems},
  year={2010}
}

@article{mihalivc2022hardware,
  title={Hardware-in-the-loop simulations: A historical overview of engineering challenges},
  author={Mihali{\v{c}}, Franc and Trunti{\v{c}}, Mitja and Hren, Alenka},
  journal={Electronics},
  volume={11},
  number={15},
  pages={2462},
  year={2022},
  publisher={MDPI}
}

@article{tao2019digital,
  title={Digital twins and cyber--physical systems toward smart manufacturing and industry 4.0: Correlation and comparison},
  author={Tao, Fei and Qi, Qinglin and Wang, Lihui and Nee, AYC},
  journal={Engineering},
  volume={5},
  number={4},
  pages={653--661},
  year={2019},
  publisher={Elsevier}
}

@inproceedings{bacic2005hardware,
  title={On hardware-in-the-loop simulation},
  author={Bacic, Marko},
  booktitle={Proceedings of the 44th IEEE Conference on Decision and Control},
  pages={3194--3198},
  year={2005},
  organization={IEEE}
}

@article{liu2021digital,
  title={Digital twin-based designing of the configuration, motion, control, and optimization model of a flow-type smart manufacturing system},
  author={Liu, Qiang and Leng, Jiewu and Yan, Douxi and Zhang, Ding and Wei, Lijun and Yu, Ailin and Zhao, Rongli and Zhang, Hao and Chen, Xin},
  journal={Journal of Manufacturing Systems},
  volume={58},
  pages={52--64},
  year={2021},
  publisher={Elsevier}
}

@inproceedings{gugerty2006case,
  title={Case study comparison of serial and ethernet digital communications technologies for transfer of relay quantities},
  author={Gugerty, Mike and Jenkins, Robin and Dolezilek, Dave},
  booktitle={proceedings of the 33rd Annual Western Protective Relay Conference, Spokane, WA},
  year={2006}
}

@misc{wiringpi_github,
  author       = {{WiringPi}},
  title        = {WiringPi: GPIO Interface Library for Raspberry Pi},
  year         = {2026},
  howpublished = {\url{https://github.com/WiringPi/WiringPi}},
  note         = {Accessed: 2026-04-09}
}

@misc{ni_compactrio,
  author       = {{NI}},
  title        = {CompactRIO Systems (cRIO)},
  year         = {2026},
  howpublished = {\url{https://www.ni.com/en/shop/compactrio.html}},
  note         = {Accessed: 2026-04-09}
}

@misc{ni_linux_rt,
  author       = {{NI}},
  title        = {Introduction to NI Linux Real-Time},
  year         = {2025},
  howpublished = {\url{https://www.ni.com/en/shop/linux/introduction-to-ni-linux-real-time.html}},
  note         = {Accessed: 2026-04-09}
}

@misc{ni_crio_arch,
  author       = {{NI}},
  title        = {Specifications Explained: CompactDAQ and CompactRIO Chassis and Controllers},
  year         = {2025},
  howpublished = {\url{https://www.ni.com/en/support/documentation/supplemental/17/specifications-explained--cdaq-and-crio-chassis//
  -and-controllers.html}},
  note         = {Accessed: 2026-04-09}
}

@article{Zhang2026MBSECoPilot,
  author  = {Zhang, Wenheng and Cockburn, Callum and Henshaw, Michael and Douglas, Peter and Palmer, Paul and Olivier-Myall, Joshua and Ji, Siyuan},
  title   = {{MBSE} {Co-Pilot}: A Research Roadmap},
  journal = {Systems Engineering},
  volume  = {29},
  number  = {1},
  pages   = {20--33},
  year    = {2026},
  doi     = {10.1002/sys.70011}
}

@INPROCEEDINGS{Li2025LLMSysMLv2,
  author={Li, Zirui and Husung, Stephan and Wang, Haoze},
  booktitle={2025 IEEE International Symposium on Systems Engineering (ISSE)}, 
  title={LLM-Assisted Semantic Alignment and Integration in Collaborative Model-Based Systems Engineering Using SysML v2}, 
  year={2025},
  volume={},
  number={},
  pages={1-8},
  doi={10.1109/ISSE65546.2025.11369983}}

\end{document}